\documentclass[journal,12pt,onecolumn,draftclsnofoot]{IEEEtran}
\usepackage[ruled,vlined]{algorithm2e}
\usepackage{amsmath,nccmath}
\usepackage{amssymb}
\usepackage{array}
\usepackage{bbm}
\usepackage{cite}
\usepackage{color}
\usepackage{mathtools}
\usepackage{multirow}
\usepackage{setspace}
\usepackage[caption=false]{subfig}
\usepackage{floatrow}
\usepackage[hidelinks]{hyperref}
\newfloatcommand{capbtabbox}{table}[][\FBwidth]
\newcolumntype{M}[1]{>{\centering\arraybackslash}m{#1}}
\newcolumntype{P}[1]{>{\centering\arraybackslash}p{#1}}
\newenvironment{subroutine}[1][htb]{\begin{algorithm}[#1]}{\end{algorithm}}

\newtheorem{remark}{Remark}
\makeatletter
\let\sv@thm\@thm
\def\@thm{\let\indent\relax\sv@thm}
\makeatother
\restylefloat{table}
\DeclareMathOperator*{\argmax}{argmax}

\begin{document} 
\title{AI/ML based Joint Source and Channel Coding for HARQ-ACK Payload}
\author{Akash Doshi, Pinar Sen, Kirill Ivanov, Wei Yang, June Namgoong, Runxin Wang, Rachel Wang, Taesang Yoo, Jing Jiang and Tingfang Ji \thanks{The authors are with Qualcomm Technologies, Inc (e-mail: akasdosh@qti.qualcomm.com). Date of current version: Sep 25, 2025. This paper was presented in part at the $13^{\mathrm{th}}$ International Symposium on Topics in Coding, August 2025 in the Session for Coding and AI \cite{doshi2025ai}.} }

\maketitle 
\normalsize
\begin{abstract}
Channel coding from 2G to 5G has assumed the inputs bits at the physical layer to be uniformly distributed. However, hybrid automatic repeat request acknowledgement (HARQ-ACK) bits transmitted in the uplink are inherently non-uniformly distributed. For such sources, significant performance gains could be obtained by employing joint source channel coding, aided by deep learning-based techniques. In this paper, we learn a transformer-based encoder using a novel ``free-lunch'' training algorithm and propose per-codeword power shaping to exploit the source prior at the encoder whilst being robust to small changes in the HARQ-ACK distribution. Furthermore, any HARQ-ACK decoder has to achieve a low negative acknowledgement (NACK) error rate to avoid radio link failures resulting from multiple NACK errors. We develop an extension of the Neyman-Pearson test to a coded bit system with multiple information bits to achieve Unequal Error Protection of NACK over ACK bits at the decoder. Finally, we apply the proposed encoder and decoder designs to a 5G New Radio (NR) compliant uplink setup under a fading channel, describing the optimal receiver design and a low complexity coherent approximation to it. Our results demonstrate 3 -- 6 dB reduction in the average transmit power required to achieve the target error rates compared to the NR baseline, while also achieving a 2 -- 3 dB reduction in the maximum transmit power, thus providing for significant coverage gains and power savings.
\end{abstract}


\section{Introduction}
\subsection{Motivation} \label{subsec:motivation}
Wireless systems from 2G to 5G are designed based on Shannon's Separation Theorem \cite{csisz1981ar}. While design of channel coding at the physical (PHY) layer assumes a uniform source prior, the design of source coding ignores the channel statistics. Moreover, in practical communication systems, channel coding and source coding for the data channel work at different layers of the protocol stack. As we asymptotically approach the Shannon limit on both source and channel coding, we begin to reconsider the uniform source assumption that was used to design channel coding at the PHY layer. 

Some sources/messages communicated at PHY may be inherently non-uniform, such as the hybrid automatic repeat request acknowledgement (HARQ-ACK) bits transmitted in the uplink (UL). HARQ-ACK bits are generated in response to successful decoding (Cyclic Redundancy Check (CRC) pass) of Physical Data Shared Channel (PDSCH) packets received in the downlink (DL). In a typical gNB\footnote{A base station is referred to as gNB in 3GPP terminology.} implementation, outer loop link adaptation (OLLA) is used to ensure that the PDSCH block error rate (BLER) remains below a target (e.g.,10\%), implying HARQ-ACK bits are ACK with probability $ \geq 0.9$. As a result, the ACK to NACK ratio would be approximately 9 to 1. In spite of the biased source prior, no source coding or compression has been applied to HARQ-ACK bits in the current standards, thus presenting a significant opportunity for performance improvement in 6G. 

Furthermore, since HARQ-ACK bits are ACK with probability $\sim 0.9$, maximum a-posteriori (MAP) decoding would lead to ACK being protected more than NACK. While an ACK error at the gNB would lead to an unnecessary PDSCH retransmission, a NACK error would go undetected at the PHY layer and only be detected at the application (APP) layer, with $\geq 8$ NACK errors leading to Radio Link Failure (RLF)\footnote{Refer \textit{maxULRLCRetransmissions} parameter in \cite{3gpp.38.133}.}. Hence it is of greater importance to have a low NACK error rate than a low ACK error rate -- this is referred to as Unequal Error Protection (UEP) of source bits. For instance, \cite{3gpp.38.104} dictates that for Physical Uplink Control Channel (PUCCH) Format 1, the NACK error rate should not exceed 0.1\% at the specified signal-to-noise ratio (SNR), while the ACK error rate should not exceed 1\%. However, HARQ-ACK decoders assume the source bits are i.i.d. and uniformly distributed and do not take UEP into account. Consequently, the UE has to transmit at a power high enough to sustain both a NACK and ACK error rate of $\leq 0.1$ \%. An optimal decoder should utilize the higher target ACK error rate to reduce the average transmit power and/or improve uplink coverage.

AI/ML is a powerful tool that one could utilize to design joint source and channel coding (JSCC) schemes for PHY layer communications. However, AI/ML based solutions, being data-driven, are typically scenario specific and hence not robust to slight changes in the underlying data distribution \cite{marcus2018deep, welling2019we}. While the algorithmic solutions that drive channel estimation, demodulation and decoding in modern cellular modems are analytical and data/distribution agnostic, AI/ML solutions fundamentally look to exploit patterns in the data distribution instead of designing an algorithmic solution to the problem at hand. For an AI/ML based JSCC scheme to be considered an effective solution, one must be able to demonstrate its robustness to distribution shifts and practical impairments.

\subsection{High Level Problem Statement} \label{subsec:problem_statement}
Suppose $k$ HARQ-ACK bits $b_0,b_1, \ldots b_{k-1}$ are encoded to $n$ coded symbols and transmitted, with each message sequence $m \in [2^k]$ occurring with probability $\pi_m$. A bit value of $1$ represents ACK and $0$ represents NACK. Denote the received signal by $\mathbf{y}$. Our objective is to learn a codebook $\mathcal{C}_u = \{\mathbf{c}_{u,m}\}_{m \in [2^k]}$ and design a decoding algorithm $\mathcal{D}(\mathbf{y};\mathcal{C}_u,\{\pi_m\}_m)$ that together minimize $\mathrm{SNR}_{\mathrm{th}} = \max(\mathrm{SNR}_{\mathrm{ack}}, \mathrm{SNR}_{\mathrm{nack}})$ where $\mathrm{SNR}_{\mathrm{ack}}$ is the minimum SNR where the $p(\hat{b}_i \neq b_i | b_i = 1)  \leq 1$\% and $\mathrm{SNR}_{\mathrm{nack}}$ is the minimum SNR where the $p(\hat{b}_i \neq b_i | b_i = 0) \leq 0.1$\%. We do not constrain the codewords of $\mathcal{C}_u$ to have the same power. Instead, we look to learn the optimal power to be assigned to each codeword by ensuring that $\sum_m \pi_m (\|\mathbf{c}_{u,m}\|^2_2/n) = 0.5$. For notational convenience, we will separate $\mathcal{C}_u$ into the normalized codebook $\mathcal{C}$ and the codeword powers $\mathbf{P}$ in the remainder of the paper.

\subsection{Related Work} \label{subsec:related_work}
Designing practical JSCC schemes for PHY layer communications is challenging. The NR channel code for short block length ($3 \leq k \leq 11$ where $k$ is the number of information bits) is a linear code with a pre-specified generator matrix \cite{3gpp.38.212}. For this regime, maximum likelihood (ML) decoding is used in standard evaluations to compute bit/block error rates (BER/BLER). For $k > 11$, Polar codes are utilized and ML decoding becomes computationally unfeasible. Typically, HARQ-ACK transmission falls into the short block length regime with the maximum value of $k = 11$ being achieved only in the case of carrier aggregation. As with other NR channel codes, the generator matrix design was optimized assuming a uniform prior on the information bits.

A recent survey \cite{gunduz2024joint} provides a very comprehensive overview of classical approaches to practical code design for JSCC. Most of the works surveyed \cite{hagenauer1995source, murad1998joint, jeanne2005joint} apply well-known source coding techniques such as Huffman\cite{huffman1952method} and Lempel-Ziv\cite{ziv1977universal} and modify trellis-based decoders such as Viterbi to perform joint-source channel decoding. However, this involves using variable-length source codes which would make the channel code rate content dependent. Meanwhile, JSCC using LDPC codes and Polar codes for both source and channel coding has been investigated in \cite{fresia2009optimized} and \cite{dong2021joint} respectively. In addition to the source coding having a variable rate, they also require highly biased source priors to yield performance benefits. 

Recently, the temporal correlation among DCI grants has been exploited in \cite{liu2024lossless} to perform source coding via Arithmetic/Huffman coding. However, they did not jointly optimize the channel code. In parallel, lossless compression techniques have employed machine learning \cite{goyal2018deepzip} to exploit the underlying distribution of source data and achieve lower compression ratio compared to traditional schemes. Since these techniques were not proposed in the context of PHY layer control signaling, they did not consider jointly optimizing the channel code as well. To the best of the authors knowledge, there is no existing work that looks to exploit the non-uniform source distribution of DL HARQ-ACK when transmitted in the uplink.

\subsection{Contributions} \label{subsec:contributions}
In this paper, we make several key contributions to encoder and decoder design for the PUCCH HARQ-ACK payload in the short block length regime. Since this is a relatively unexplored field as outlined in Section \ref{subsec:related_work}, this paper makes several fundamental contributions to PUCCH HARQ-ACK transmission and reception design, including non-linear transformer-aided codebook design and per-codeword power shaping to exploit the HARQ-ACK source prior, a bitwise maximum a-posteriori (MAP) based receiver with a custom threshold to protect NACK bits more than ACK bits and a practical receiver design framework that provides for power shaping of reference signals whilst performing channel estimation as elaborated below.

\textbf{AI/ML based encoder design:} We outline a transformer-based non-linear codebook design to exploit the non-uniform HARQ-ACK source prior, highlighting how the absence of positional embeddings reduces the size of the lookup table (LUT) used to store the codebook. We also describe how the output of the transformer can be quantized to the binary field during training. As part of the simulation details, we also present the training loss function, its theoretical underpinnings and detail the ``free-lunch'' training algorithm - training a common codebook for a range of source priors.

\textbf{Power Shaping:} We present a novel alternative to variable length source coding - per-codeword power shaping. By assigning less power to more frequent codewords and vice versa, we demonstrate how the average SNR required to achieve a target error rate is minimized. Further, we describe power shaping techniques that are more robust to changes in the source distribution and/or require less signaling overhead.

\textbf{Bitwise UEP:} In order to protect the NACK bits more than ACK bits, we formulate a decoder that involves comparing the ratio of the bitwise likelihoods to a custom threshold. Further, we develop an extension inspired by the generalized likelihood ratio test from estimation theory to overcome partial knowledge of the source prior at the receiver and propose an approximation to it that significantly reduces computational complexity.

\textbf{Practical Receiver Design:} Making minimal assumptions with regards to the channel model, we first describe the optimal approach to computing the bitwise likelihoods when transmitting over a fading channel assuming a noncoherent receiver. We then factorize the likelihoods to separate the channel estimation and decoding steps to develop a coherent receiver, and finally describe how the coherent receiver processing can be significantly simplified in certain scenarios to achieve computational complexity on par with a traditional maximum likelihood (ML) decoder. 

\subsection{Notation \& Organization} \label{subsec:notation}
We use bold uppercase $\mathbf{A}$ to denote a matrix and bold lowercase $\mathbf{a}$ to denote a vector, unless stated otherwise. Similarly, $\mathbf{1}$ denotes the all-ones vector. A matrix $\mathbf{A}$ can be obtained by stacking column vectors as $\mathbf{A} = [\mathbf{a}_0, \mathbf{a}_1 \ldots \mathbf{a}_{N - 1}]$. $\mathbf{A}[i,j]$ represents the element of $\mathbf{A}$ located at the $i^{\mathrm{th}}$ row and $j^{\mathrm{th}}$ column. For vectors $\mathbf{a}$ and $\mathbf{b}$ of length $n$, $\mathbf{a} \odot \mathbf{b}$ represents the elementwise multiplication of the two vectors. $\|\mathbf{a}\|_2$ denotes the Frobenius norm of $\mathbf{a}$. In this paper, we will often invoke the matrix inversion lemma in its simplified form, which states that
\begin{equation} \label{eq:inv_lemma_simple}
    (\mathbf{I} + \mathbf{UV})^{-1} = \mathbf{I} - \mathbf{U}(\mathbf{I} + \mathbf{VU})^{-1} \mathbf{V}
\end{equation}
for conformable matrices $\mathbf{U}$ and $\mathbf{V}$. Given a scalar value $n$, $n! = n(n-1)\ldots1$.

The paper is organized as follows. A frequency domain system model is outlined in Section \ref{sec:system_model}, followed by a detailed description of the encoder design and the concept of power shaping in Section \ref{sec:enc_design} and \ref{sec:ps} respectively. We then outline the receiver design for an AWGN channel that assumes perfect channel estimation in Section \ref{sec:awgn_receiver}, and then elaborate on practical considerations under a NR compliant fading channel in Section \ref{sec:practical_receiver}. We present the simulation details common to both the AWGN and practical receiver design in Section \ref{sec:simulation_details}, before presenting the results for the AWGN and practical receivers, along with their specific simulation details, in Section \ref{sec:awgn_receiver_results} and \ref{sec:practical_receiver_results} respectively. Finally, we summarize the paper and describe potential future directions in Section \ref{sec:conclusions}.

\section{System Model} \label{sec:system_model}
Consider a channel code mapping $k$ information bits to $n$ coded symbols, henceforth identified by the tuple $(k,n)$. An information bit value of 0 corresponds to NACK, while 1 corresponds to ACK. A message sequence $\mathbf{b}_m = \{b_0,b_1, \ldots b_{k-1}\} \in \{0,1\}^k$ consisting of $k$ information bits is jointly encoded and modulated to $n$ real values (or equivalently $n/2$ complex values). The message sequence index is denoted by $m$ (for e.g. if $k = 4$, then the message sequence 0101 corresponds to index 5). The complex valued encoded sequence is denoted by $\mathbf{c}_m$. Each codeword $\mathbf{c}_m$ has norm $\|\mathbf{c}_m\|^2_2 = n/2$ and is associated with a prior probability $\pi_m$ such that $\sum_{m \in [2^k]} \pi_m = 1$. If the encoding is binary, the intermediate encoded bit sequence is denoted by $\mathbf{e}(\mathbf{b}_m)$. Denote the codebook by $\mathcal{C} = \{\mathbf{c}_0, \mathbf{c}_1, \ldots \mathbf{c}_{2^k-1}\}$. Subsequently, each codeword $\mathbf{c}_m$ is multiplied by a scalar per-codeword power shaping factor $\sqrt{\alpha_m}$ (to be introduced in Section \ref{sec:ps}) such that $\sum_m \pi_m \alpha_m = 1$. In current NR specifications \cite{3gpp.38.212}, all codewords are transmitted with equal power i.e. $\alpha_m = 1$. 

\subsection{AWGN setup} \label{subsec:awgn_system_model}
The coded sequence $\sqrt{\alpha_m}\mathbf{c}_m$ is transmitted over a SISO AWGN channel such that
\begin{equation}
\mathbf{y} = g\sqrt{\alpha_m}\mathbf{c}_m + \mathbf{n}
\end{equation}
where $\mathbf{n} \sim \mathcal{CN}(0,\sigma^2\mathbf{I}_{n/2})$ and $\mathrm{SNR} = \sum_m \pi_m \alpha_m g^2/\sigma^2 = g^2/\sigma^2$.

\subsection{NR-compliant fading channel setup} \label{subsec:fading_system_model}
\begin{figure}
    \centering
    \includegraphics[width = 6.5in]{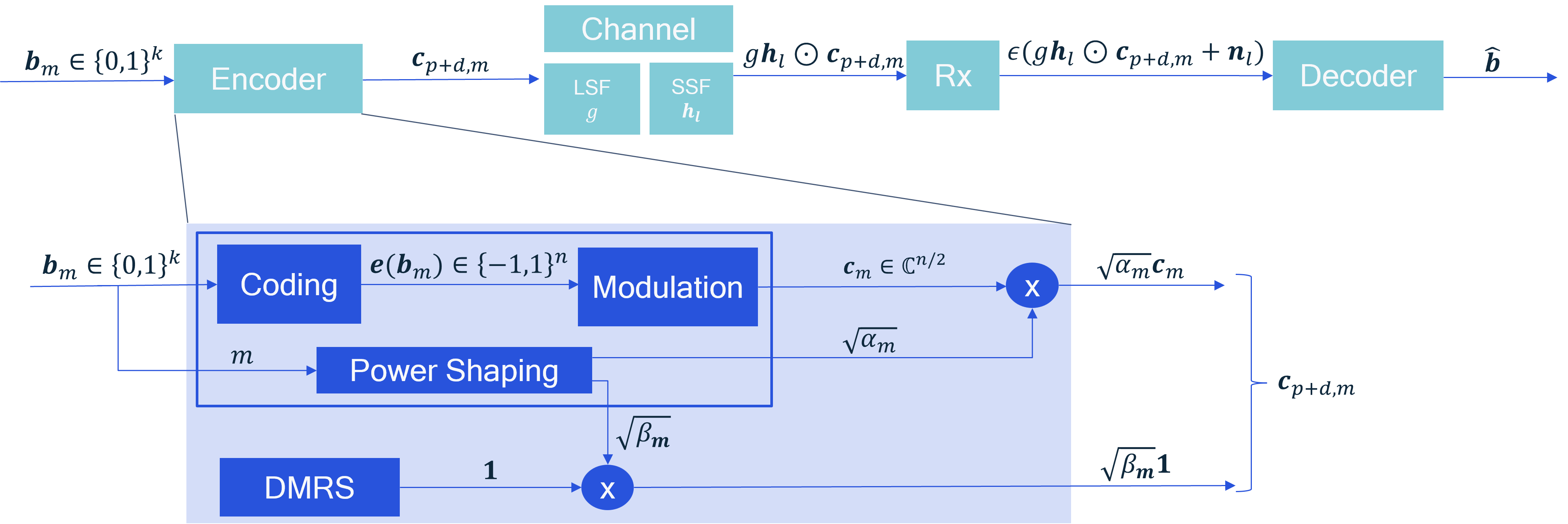}
    \caption{E2E FD pipeline for uplink SIMO PUCCH HARQ-ACK transmission per Rx antenna.}
    \label{fig:e2e}
\end{figure}

Consider the end-to-end (E2E) frequency domain (FD) pipeline for uplink transmission of PUCCH\footnote{Note that a similar formulation can be derived for HARQ on PUSCH as well. We focus only on PUCCH for simplicity of the SIMO setup.} HARQ-ACK as depicted in Fig. \ref{fig:e2e}. We assume a SIMO setup with $N_r$ receive antennas, since PUCCH is transmitted from a single antenna port \cite{johnson20195g}. Under the assumption of coherent reception i.e. excluding PUCCH Format 0, we would also transmit $N_p$ Demodulation Reference Symbols (DMRS) resource elements (REs) which we set to 1 w.l.o.g. given a single transmit port. DMRS REs are utilized by the receiver to perform channel estimation (ChEST). Suppose that the DMRS REs are also multiplied by a power shaping factor $\sqrt{\beta_m}$. In this paper, we will be considering two cases - no power shaping on the DMRS i.e. $\beta_m = 1$ and data dependent DMRS power shaping i.e. $\beta_m = \alpha_m$. In either case, observe that $\sum_m \pi_m \beta_m = 1$.

Denote by $\mathbf{c}_{p+d,m} = [\sqrt{\beta_m} \mathbf{1}, \sqrt{\alpha_m} \mathbf{c}_m]$ the OFDM resource grid containing the DMRS and data REs. This comprises the FD transmit signal consisting of $N = N_p + N_d$ REs, where $N_d = n/2$, which is to be sent over the FD channel $\mathbf{H} \in \mathbb{C}^{N \times N_r}$. The channel $\mathbf{H}$ can be expressed as $g[\mathbf{h}_0, \mathbf{h}_1, \ldots \mathbf{h}_{N_r}]$ where $g$ captures the Tx power and path loss while $\mathbf{h}_l \in \mathbb{C}^N$ is the small scale fading channel corresponding to $l^{\mathrm{th}}$ receive antenna such that $\mathbb{E}[\|\mathbf{h}_l[j]\|^2] = 1$ $\forall j \in [N]$. Furthermore, we can divide $\mathbf{h}_l$ into the channel on the pilot and data REs as $\mathbf{h}_l = [\mathbf{h}_{p,l}, \mathbf{h}_{d,l}]$ such that $\mathbf{h}_l \odot \mathbf{c}_{p+d,m} = [\sqrt{\beta_m} (\mathbf{h}_{p,l}  \odot \mathbf{1}), \sqrt{\alpha_m} (\mathbf{h}_{d,l} \odot \mathbf{c}_m)]$. The received signal on the $l^{\mathrm{th}}$ receive antenna is
\begin{equation} \label{eq:received_signal}
    \mathbf{y}_l = g\mathbf{h}_l\odot \mathbf{c}_{p+d,m} + \mathbf{n}_l,
\end{equation}
where $\mathbf{n}_l \sim \mathcal{CN}(\mathbf{0},\sigma^2 \mathbf{I})$\footnote{The system modeling can be extended w.l.o.g. to provide for any receiver scaling that is common to both signal and noise.}.

Denote the received signal (across all receive antennas) by $\mathbf{Y} = [\mathrm{y}_1, \ldots \mathrm{y}_{N_r}]$. Depending on whether we choose to perform coherent or noncoherent decoding, the received signal is passed to a channel estimator module before being input to a decoder. Furthermore, we can define the long term average SNR (per RE and per receive antenna) and noise variance (NV) measured on the DMRS tones as 
\begin{equation}
\begin{aligned}
    \mathrm{SNR} &=  \frac{\sum_m \pi_m \beta_m g^2}{\sigma^2} = \frac{g^2}{\sigma^2}\\
    \mathrm{NV} &= \sigma^2.
\end{aligned}
\end{equation}
Our objective is to design a codebook $\mathcal{C}$, power shaping vector $\mathbf{P}$ and a decoder $\mathcal{D}$ that can exploit the HARQ-ACK source prior and hence reduce the average transmit power required to preferentially protect NACK over ACK bits. We will introduce the encoder design aspects in Section \ref{sec:enc_design} and \ref{sec:ps}, the receiver design in Section \ref{sec:awgn_receiver} and \ref{sec:practical_receiver} and the UEP performance criteria in Section \ref{subsec:uep_criterion}.

\section{Encoder Design} \label{sec:enc_design}
\subsection{Transformer (TF) Architecture}
\begin{figure}
    \centering
    \includegraphics[width = 3.1in]{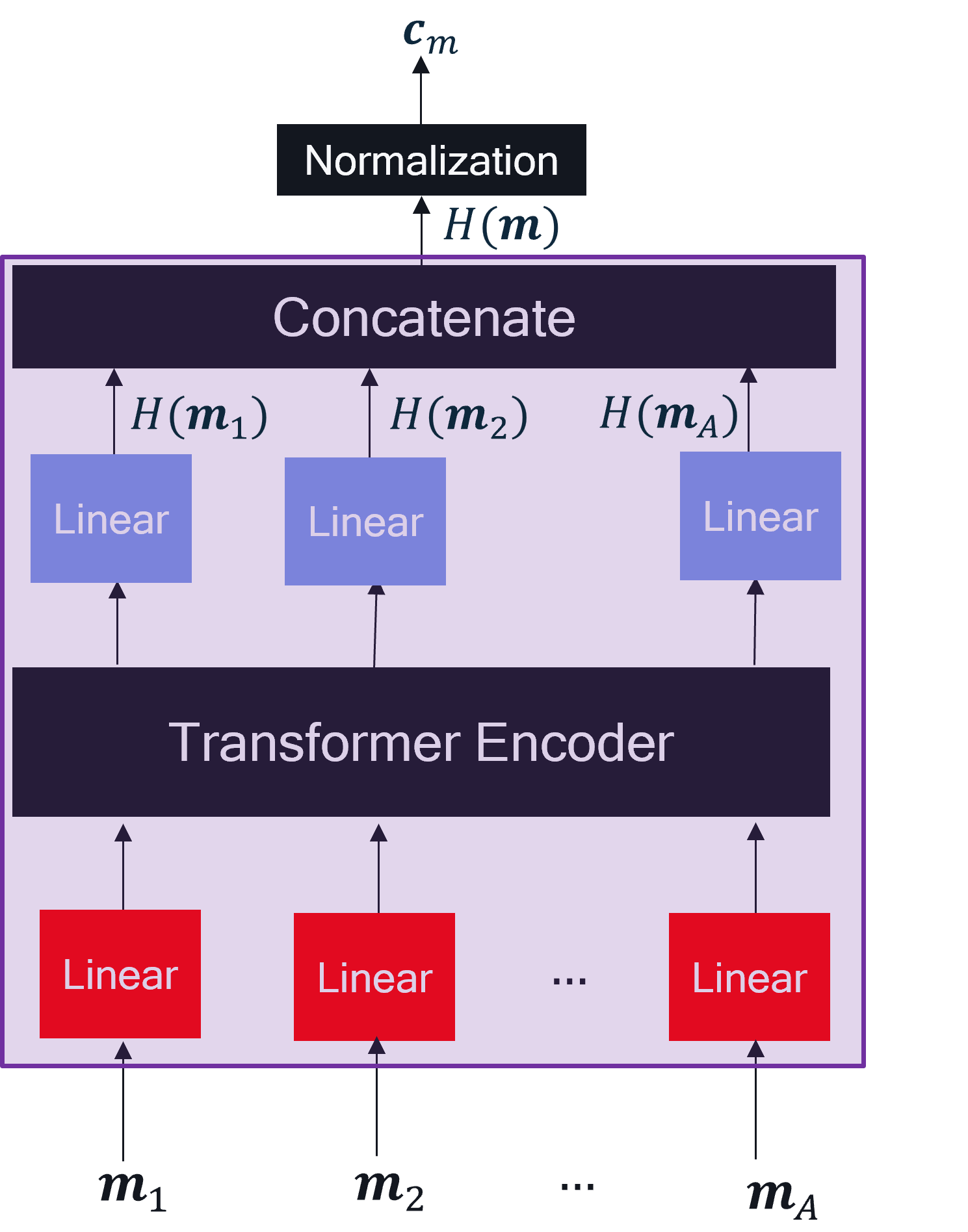}
    \caption{Transformer-based Architecture for encoding PUCCH HARQ-ACK}
    \label{fig:transformer}
\end{figure}

We utilize a transformer (TF) encoder \cite{vaswani2017attention} to encode the input bits. The architecture is illustrated in Fig. \ref{fig:transformer}. The input bit sequence $\mathbf{b}_m \in \{0,1\}^k$ is divided into $A$ tokens of size $\lceil k/A \rceil$. If $k < A\lceil k/A \rceil$, the first $k$ positions are filled with input bits and the remaining $A\lceil k/A \rceil - k$ positions are set to 0. Subsequently, we BPSK modulate the input bits $\mathbf{m} = (-1)^{\mathbf{b}_m}$ before inputting the tokens to the transformer. The transformer encoder consists of $L = 6$ layers. Each layer consists of the standard self attention and feedforward networks with residual connections. We utilize an embedding length\footnote{Refer https://pytorch.org/docs/stable/generated/torch.nn.TransformerEncoder-Layer.html for meaning of each hyperparameter} of $\mathrm{d}_\mathrm{model} = 128$, feedforward dimension $\mathrm{dim}_\mathrm{feedforward}=512$ and $\mathrm{nhead}=8$. The tokens input to the transformer and output by it are passed through Layer Norm \cite{xiong2020layer} normalization. Positional embedding is not added to the input tokens. The transformer outputs $A$ tokens. Each output token is passed through a linear layer $H$ to obtain $\lceil n/A \rceil$ coded symbols per token. The coded symbols computed from the output tokens are concatenated to obtain the codeword of $n$ coded symbols, which can then be normalized to ensure the total codeword power of each codeword is $n/2$. Finally, the transformer output can be reshaped to obtain $\mathbf{c}_m$, an $n/2$ length complex codeword.

\subsection{Codebook Properties} \label{subsec:codebook_properties}
Since the encoding operations involve computing self-attention on the message tokens, the codebook is non-linear. Non-linearity implies that each codeword will have a different Block Error Rate (BLER)-SNR curve, leading to a per-codeword BLER spread.

Since we do not utilize positional embedding, the codebook learnt is group permutation symmetric with a group size of $\lceil k/A \rceil$. This implies we do not need to store the embedding of any group permutations of the input message sequence. Given an input sequence $\mathbf{x}$, we do not need to invoke the transformer to compute the encoding of any sequence that can be written as $\mathbf{Px}$, where $\mathbf{P} \in \{0,1\}^{A \times A}$ and $\mathbf{x} = [ [\mathbf{m}_0^T], [\mathbf{m}_1^T] \ldots ,[\mathbf{m}_{A-1}^T]] \in \{-1,1\}^{A \times \lceil k/A \rceil}$. This implies that for any input sequence $\mathbf{x}$, there are at most $A! - 1$ orderings of its input tokens\footnote{If some of the tokens are identical, then the number of distinct orderings will be smaller.} whose encoding can be computed by group permutation of $H(\mathbf{x})$. This property can be exploited to reduce the size of look-up table (LUT) used to store the codebook. Furthermore, increasing the number of tokens $A$ (or equivalently reducing the token size $\lceil k/A \rceil$) increases the degree of group permutation symmetry in the codebook. In other words, larger symmetry reduces the size of the LUT required to store the codebook (refer Appendix \ref{subsec:perm_equi} for further details).

\subsection{Quantizing Transformer Output} \label{subsec:quantization}
So far the $n$ coded symbols output by the transformer are real-valued, hence we can view this as a joint source-channel coding and modulation (JSCCM) scheme. If the transformer output is unquantized, and the coded symbols were to be transmitted on a DFT-s-OFDM waveform, we can consider that the coded symbols are effectively being transmitted as consecutive time-domain symbols. In that case, unquantized symbols can lead to a large peak-to-average power ratio (PAPR). To lower the PAPR, one approach could be quantizing the output of transformer to $\{0,1\}^n$ and then applying QPSK modulation as done in NR \cite{3gpp.38.213}. In order to learn a quantized codebook, we adopt a simplified approach to VQ-VAE \cite{van2017neural} called finite scalar quantization (FSQ) \cite{mentzer2023finite}. Denoting the unquantized transformer output by $\mathbf{c}_{\mathrm{unquant}}$, we have
\begin{equation}
\mathbf{c}_{\mathrm{quant}} = \mathrm{sign}(1 - 2\sigma(\mathbf{c}_{\mathrm{unquant}})),
\end{equation} 
where $\sigma(x) = 1/(1 + e^{-x})$. Since $\sigma(x)$ constrains the output to lie in $(0,1)$, $1 - 2\sigma(x)$ constrains it to $(-1,1)$. In order to perform the quantization differentiably during training, we set
\begin{equation}
    \mathbf{c}_{\mathrm{train}} = 1 - 2\sigma(\mathbf{c}_{\mathrm{unquant}}) + \mathrm{sg}(\mathbf{c}_{\mathrm{quant}} - (1 - 2\sigma(\mathbf{c}_{\mathrm{unquant}}))),
\end{equation}
where $\mathrm{sg}$ stands for "stop gradient". Consequently, the gradient will flow through $1 - 2\sigma(\mathbf{c}_{\mathrm{unquant}})$ but the value of $\mathbf{c}_{\mathrm{train}}$ will be $\mathbf{c}_{\mathrm{quant}}$. We will utilize $\mathbf{c}_{\mathrm{train}}$ and  $\mathbf{c}_{\mathrm{quant}}$ during training and inference respectively.

\section{Power Shaping (PS)} \label{sec:ps}
NR channel code designs \cite{3gpp.38.212} assumed i.i.d. information bits with $p(b = 1) = 0.5$. One classical approach to exploit source non-uniformity is to apply source coding to the data prior to channel coding. Classical lossless source coding techniques such as entropy coding \cite{10666422} reduce the average number of bits required to transmit the same data by setting the word length to be the negative log probability of the word i.e. $l(\mathbf{c}_m) \propto - \log \pi_m$ where $l(\mathbf{c}_m)$ denotes the length of $\mathbf{c}_m$. Given that the source compression is lossless, this would reduce the average transmit power needed to achieve a given target BER.

However, such source coding techniques are inconvenient for wireless communication, since the channel coding parameters $(k,n)$ would become content dependent (typically they are only channel dependent). Hence, we propose instead to perform per-codeword power shaping. This implies that each codeword $\mathbf{c}_m$ will be associated with a scalar per-codeword power shaping parameter $\alpha_m$ while the length of each codeword is kept fixed. In this section, we will introduce three distinct power shaping techniques and highlight their pros and cons. We will present a comparative analysis of the gains obtained from the different PS schemes in Section \ref{subsec:ps_results}.

\subsection{Entropy PS}
Given a codeword $\mathbf{c}_m$ with prior $\pi_m$, set the codeword power $\alpha_{m} \propto  - \log \pi_m$, henceforth referred to as Entropy PS. While we have been unable to theoretically prove the optimality of Entropy PS in terms of minimizing the BER given an average power constraint, we verified that a learnable PS scheme, learnt as part of the transformer training, also converges to a power allocation similar to Entropy PS (training details in Appendix \ref{subsec:learnt_ps_appendix}). Fig. \ref{fig:entropy_ps} shows $\alpha_{m}$ as a function of $\pi_m$ for $k=11$ HARQ-ACK bits which are i.i.d. and ACK with $p= 0.9$ for both Entropy and learnt PS.
\begin{figure}
    \centering
    \includegraphics[width = 3.1in]{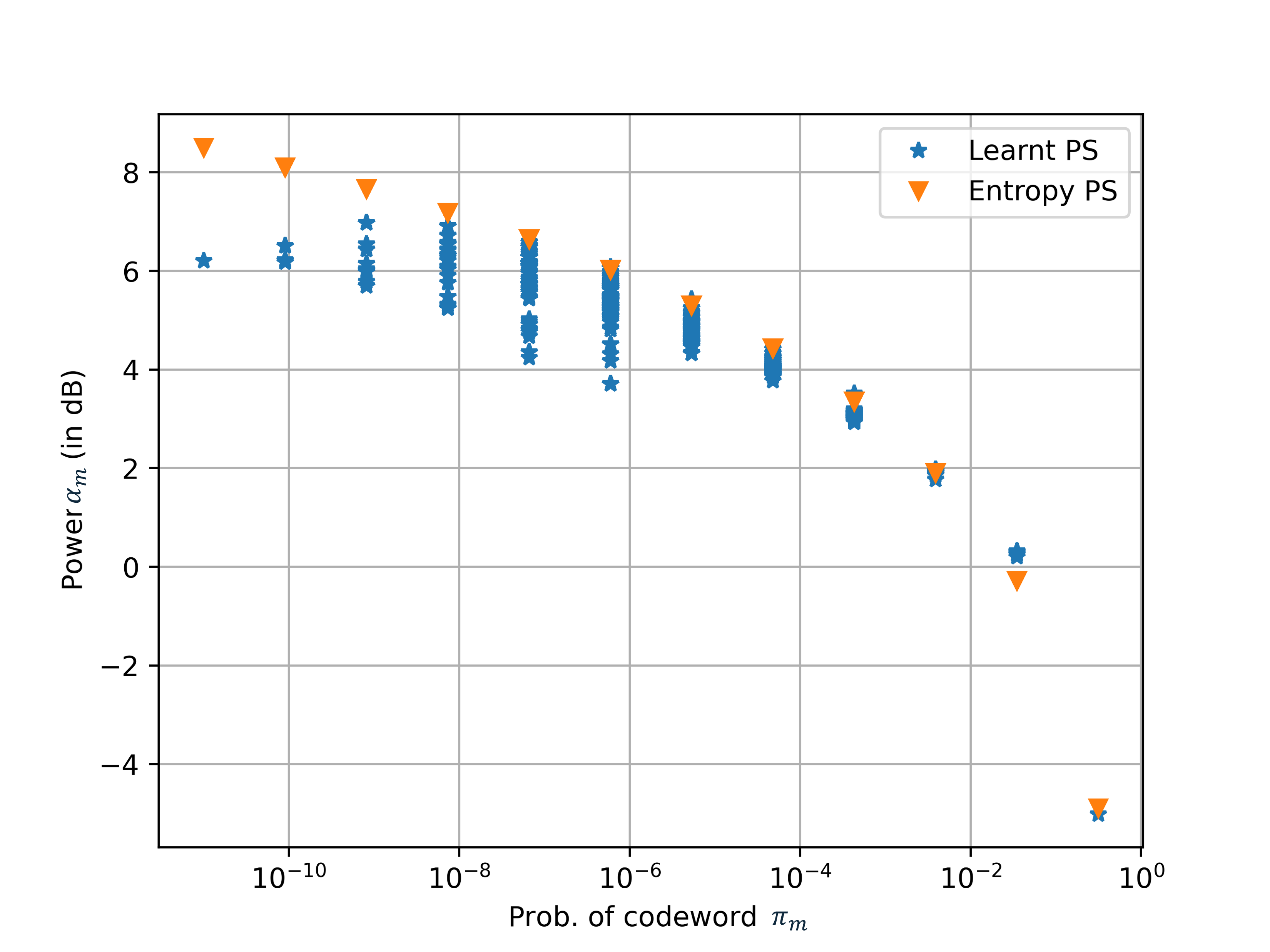}
    \caption{Entropy and Learnt PS $\alpha_m$ as a function of $\pi_m$}
    \label{fig:entropy_ps}
\end{figure}

Observe that codewords with high probability of occurrence $\pi_m$ have low power $\alpha_m$ and vice-versa. Moreover, codewords with the same probability of occurrence have the same codeword power. Since we plot as a function of $\pi_m$, Entropy PS is a single dot but the Learnt PS has a spread (if more than one codeword has the same prior probability). Since codewords with low $\pi_m$ are seen less frequently in training, this is to be expected. However, for all codewords w.p. $\geq 1e-4$, Learnt and Entropy PS are in very close agreement.

\subsection{Arithmetic PS} 
A possible drawback to Entropy PS is that any change in the prior $\mathbf{\Pi}$ changes the power shaping. We want to design a PS scheme that is more robust to changes in the prior. In other words, we want to find a representative prior $\hat{\mathbf{\Pi}}$ such that if we set the codeword power $\alpha_m \propto - \log {\hat{\pi}_m}$, it provides the highest attainable gains for all distributions under consideration subject to the constraint of a fixed PS scheme.  Parameterizing $\mathbf{\Pi}$ by $\Psi$, this would imply finding
\begin{equation} \label{eq:arithemtic_ps}
    \hat{\mathbf{\Pi}} = \arg \min_{\mathbf{\Pi}'} \mathbb{E}_{\Psi} [\mathrm{KL}(\mathbf{\Pi}(\Psi) || \mathbf{\Pi}')] = \arg \min_{\mathbf{\Pi}'} \mathbb{E}_{\Psi} \bigg[\sum_{m \in [2^k]} - \pi_{m}(\Psi) \log \pi'_m\bigg],
\end{equation}
subject to the constraint $\sum_{m} \pi'_{m} = 1$ in a Bayesian setup (with a uniform prior on $\Psi$). Utilizing Lagrange multipliers and differentiating w.r.t $\pi'_{m}$ $\forall \mathbf{c}$, we obtain
\begin{equation}
    \hat{\mathbf{\Pi}} = \mathbb{E}_{\Psi} [\mathbf{\Pi}(\Psi)].
\end{equation}
Since computing $\hat{\mathbf{\Pi}}$ involves computing the arithmetic mean over all $\mathbf{\Pi}(\Psi)$, we henceforth refer to this scheme as Arithmetic PS. Similar to Entropy PS, we will present results in Appendix \ref{subsec:learnt_ps_appendix}, demonstrating that a common learnable PS scheme, learnt as part of the transformer training, converges to a allocation similar to Arithmetic PS. 

\subsection{Step PS}
While Arithmetic PS enables utilization of a common PS scheme for a range of priors, both Entropy and Arithmetic PS lead to a large variation in power between different codewords. While power shaping was introduced to exploit the non-uniform codeword prior, we would still like to be able to control the power variation. In order to control the power variation, we propose Step PS. Step PS consists of only two power levels -- $\alpha_0 = P_0$, power of the all-ACK codeword and $\alpha_m = P_1$ $\forall m \neq 0$, power of all the other codewords.  Furthermore, Step PS can be parameterized by a single value $\delta = P_1 - P_0$, where $P_0$ and $P_1$ are in dBm and $\delta$ is in dB. By setting $P_0 < P_1$, Step PS acts as a coarse approximation to Entropy and Arithmetic PS. The power variation in Step PS can be easily reduced by lowering $\delta$, and the signaling overhead is reduced from $2^k$ codeword power values to $1$ power differential.

\vspace{2mm}
\begin{remark}
For all PS schemes, the power levels are subjected to the constraint $\sum_{m} \pi_{m}\alpha_m = 1$. This is to ensure a consistent definition of $\mathrm{SNR}$ during simulation. In other words, the long term average SNR in the presence of power shaping In practice, such a normalization would not be performed. It would instead be performed implicitly as part of the power control equation in UE behavior (refer Section 7.2.1 in \cite{3gpp.38.213}).
\end{remark}

\section{AWGN Receiver Design} \label{sec:awgn_receiver}
The encoder and power shaping designs proposed thus far seek to exploit the underlying prior of the HARQ-ACK bits to perform JSCC at the encoder. In this section, we switch our focus to enabling JSCC at the decoder while also providing for UEP of NACK over ACK by effective utilization of the codeword prior. To develop a decoder from first principles, we start with the AWGN system model detailed in Section \ref{subsec:awgn_system_model}. In this case, the codeword likelihood is given by \cite{neeser2002proper}
\begin{equation}
    p(\mathbf{y}|\mathcal{H}_m) = \frac{1}{\pi^{n/2} \mathrm{det} \mathbf{\Sigma}} \mathrm{exp}(-(\mathbf{y}-g\sqrt{\alpha_m}\mathbf{c}_m)^H \mathbf{\Sigma}^{-1}  (\mathbf{y}-g\sqrt{\alpha_m}\mathbf{c}_m))
\end{equation}
where $\mathbf{\Sigma} = \sigma^2 \mathbf{I}_{n/2}$ and $\mathcal{H}_m = \{\mathbf{c}_m, \alpha_m\}$ denotes the $m^{\mathrm{th}}$ codeword hypothesis i.e. the codeword is $\mathbf{c}_m$ and the power shaping is $\alpha_m$. 

In order to utilize the codeword prior $\mathbf{\Pi}$ at the decoder, the standard approach is to perform Maximum a-posteriori (MAP) decoding. At the codeword level, this involves identifying the codeword index $m^*$ such that
\begin{equation}
    m^* = \argmax_{m} p(\mathbf{y}|\mathcal{H}_m) \pi_m.
\end{equation}
Given our focus on preferentially protecting NACK over ACK, we pivot to bitwise decoding in place of codeword decoding. Hence, we would instead perform bitwise MAP decoding, such that
\begin{equation} \label{eq:bitwise_MAP}
\begin{gathered}
    \hat{b}_i = \argmax_{j \in \{0,1\}} p(b_i = j|\mathbf{y}) \\
   p(b_i = j|\mathbf{y}) = \frac{\sum_{m \in \mathcal{C}_{b_i = j}} p(\mathbf{y}|\mathcal{H}_m) \pi_m}{p(\mathbf{y})},
\end{gathered} 
\end{equation}
where $\mathcal{C}_{b_i = j}$ denotes the set of all codewords for which $b_i = j$. 

\subsection{Bitwise Unequal Error Protection (Bitwise UEP)}\label{subsec:bitwise_uep}
Rewriting the bitwise MAP decoding in \eqref{eq:bitwise_MAP} as a thresholding rule, we decode $\hat{b}_i = 0$ if
\begin{equation}
    \frac{\sum_{m \in \mathcal{C}_{b_i = 0}} p(\mathbf{y}|\mathcal{H}_m) \pi_m}{\sum_{m \in \mathcal{C}_{b_i = 1}} p(\mathbf{y}|\mathcal{H}_m) \pi_m} = \frac{p(\mathbf{y},b_i=0)}{p(\mathbf{y},b_i=1)} \geq 1,
\end{equation}
which further can be rewritten in terms of the bitwise likelihood as
\begin{equation} \label{eq:bit_likelihood_ratio}
    \frac{p(\mathbf{y}|b_i=0)}{p(\mathbf{y}|b_i=1)} \geq \frac{p}{1-p}.
\end{equation}
Now suppose that $p = 0.9$. Observe that $p(\mathbf{y}|b_i=0)$ has to be $0.9/(1-0.9) = 9$ times larger than $p(\mathbf{y}|b_i=1)$ for the bit to be declared NACK. Consequently, a bitwise MAP decoder would preferentially protect ACK over NACK. As outlined in Section \ref{subsec:motivation}, we need to preferentially protect NACK over ACK. This would suggest the use of a ``fake" prior $\beta$ 
\begin{equation} \label{eq:rule_beta}
    \frac{p(\mathbf{y}|b_i=0)}{p(\mathbf{y}|b_i=1)} \geq \frac{\beta}{1-\beta} \Leftrightarrow \frac{p(b_i=0|\mathbf{y})}{p(b_i=1||\mathbf{y})} \geq \frac{\beta (1-p)}{(1-\beta)p}
\end{equation}
which could be swept to achieve the desired degree of NACK over ACK protection. We will henceforth refer to this decoder as Bitwise UEP. Bitwise MAP is a special case of Bitwise UEP where $\beta/(1-\beta) = p/(1-p)$. From detection theory \cite{steven1993fundamentals}, we observe that the likelihood-based criterion in \eqref{eq:rule_beta} resembles the classical Neyman-Pearson (NP) test. However, the NP test does not require any knowledge of the prior $\mathbf{\Pi}$ since it is applicable for one bit only i.e. $k = 1$. In the case of repetition coding for $k=1$, \eqref{eq:rule_beta} simply collapses to comparing $\Re\{\mathbf{1}^H\mathbf{y}\}$ with a threshold. Depending on the target NACK error rate and the SNR, the threshold can easily be computed (refer Appendix \ref{subsec:one_bit_uep}).

Extending the NP test to $k > 1$ requires computing the posterior bit probability $p(b_i = 0|\mathbf{y})$ via Bayes rule as outlined in \eqref{eq:bitwise_MAP}. However, it is unclear how to determine $\beta$ to achieve the desired degree of NACK over ACK protection. We propose an empirical rule to determine the threshold $\beta$ (experimental evidence is presented in Appendix \ref{subsec:varying_beta}). Suppose we wish to target an ACK and NACK BER of $\mathrm{BER}_{\mathrm{ACK,t}}$ and $\mathrm{BER}_{\mathrm{NACK,t}}$ respectively. Then, set
\begin{equation} \label{eq:thresh}
    \frac{\beta}{1-\beta} \approx \frac{\mathrm{BER}_{\mathrm{NACK,t}}}{\mathrm{BER}_{\mathrm{ACK,t}}}.
\end{equation}
Suppose $\mathrm{BER}_{\mathrm{ACK,t}} = 1$\% and $\mathrm{BER}_{\mathrm{NACK,t}} = 0.1$\%. Then $\beta/(1- \beta) = 0.1$ and $p(\mathbf{y}|b_i=1)$ has to be $10$ times larger than $p(\mathbf{y}|b_i=0)$ for $b_i$ to be declared ACK, thus preferentially protecting NACK. 

\subsection{Max Log MAP Generalized Likelihood Ratio Test (MLM GLRT)} \label{subsec:mlm_glrt}
One caveat to performing Bitwise UEP is the requirement of perfect knowledge of the prior $\mathbf{\Pi} = \{\pi_m\}_m$, which may not hold in reality. For instance, since OLLA ensures PDSCH BLER is $\leq 10$\%, the BS may know $p = 0.9$. However, owing to downlink channel statistics, consecutive HARQ-ACK bits may be correlated and the BS may not be privy to the correlation value. 

Consider a parameterized representation of $\mathbf{\Pi} = f(\Psi_0, \Psi_1)$. In other words, given $\Psi_0$ and $\Psi_1$, we can compute each of the $2^k$ codeword probabilities. Further, assume that $\Psi_0$ is known and $\Psi_1$ is unknown. The generalized likelihood ratio test (GLRT) \cite{steven1993fundamentals} replaces the unknown parameters under each hypothesis by their maximum likelihood estimates under that hypothesis. Given $\Psi_1$ is unknown, GLRT can be applied to \eqref{eq:rule_beta} to obtain
\begin{equation} \label{eq:glrt}
    \frac{\max_{\Psi_1} p_{\Psi_1}(\mathbf{y}|b_i=0)}{\max_{\Psi_1} p_{\Psi_1}p(\mathbf{y}|b_i=1)} \geq \frac{\beta}{1-\beta}.
\end{equation}
From \eqref{eq:bit_likelihood_ratio}, it follows that $p_{\Psi_1}(\mathbf{y}|b_i=j) = \sum_{m \in \mathcal{C}_{b_i = j}} p(\mathbf{y}|\mathcal{H}_m) \pi_{m}(\Psi_1) / p(b_i = j)$\footnote{Observe that this does not capture the dependence of the marginal probability $p(b_i = j)$ on $\Psi_1$. This is valid only if $p(b_i = j)$ is independent of $\Psi_1$ as will be considered in Section \ref{subsec:prob_model}.}.

Furthermore, in order to reduce computation complexity, we observe that $p(\mathbf{y},b_i = j)$ is computed as a weighted sum of exponentials $\sum_{m \in \mathcal{C}_{b_i = j}} p(\mathbf{y}|\mathcal{H}_m) \pi_{m}$. Approximating a sum of exponentials by the dominating exponential and taking $\mathrm{log}$ on both sides of \eqref{eq:glrt}, we obtain the decision rule 
\begin{equation} \label{eq:mlm_glrt}
    \max_{\Psi_1} \max_{m \in \mathcal{C}_{b_i = 1}} \mathcal{L}(m,\Psi_1) - \max_{\Psi_1} \max_{m \in \mathcal{C}_{b_i = 0}} \mathcal{L}(m,\Psi_1) \gtrless \log{\frac{\beta(1-p)}{p(1-\beta)}}
\end{equation}
where
\begin{equation}
    \mathcal{L}(m,\Psi_1) =  -(\mathbf{y}-g\sqrt{\alpha_m}\mathbf{c}_m)^H \mathbf{\Sigma}^{-1}  (\mathbf{y}-g\sqrt{\alpha_m}\mathbf{c}_m) + \log {\pi_m}(\Psi_1).
\end{equation}
We refer to this approximation as the Max Log MAP (MLM) approach \cite{hagenauer1994iterative}. MLM GLRT provides two benefits - the ability to compute the bitwise likelihood with partial knowledge of the codeword priors and a reduced complexity implementation (compared to Bitwise UEP). Since we compute an approximation to $p(\mathbf{y}|b_i = 0)$, the value of $\beta$ as dictated by \eqref{eq:thresh} may no longer be exactly optimal i.e. achieve the target ACK and NACK error rates at the same SNR. However, in the absence of \eqref{eq:thresh}, an optimal value for $\beta$ cannot be determined in practice since the gNB is unable to ascertain if the received signal has been correctly decoded owing to a lack of CRC in the short block-length regime. Hence, to maintain a fair comparison, we will not change $\beta$ for MLM GLRT when comparing it to Bitwise UEP in Section \ref{subsec:mlm_glrt_results}.

\section{Fading Channel Receiver Design} \label{sec:practical_receiver}
Having designed a Bitwise UEP receiver from first principles assuming a SISO AWGN channel with perfect channel estimation in Section \ref{sec:awgn_receiver}, we now tackle the fading channel setup outlined in Section \ref{subsec:fading_system_model}. In order to extend the receiver design to a SIMO fading channel, we recall that the SIMO received signal is $\mathbf{Y} = [\mathbf{y}_1, \ldots \mathbf{y}_{N_r}]$, where $\mathbf{y}_l$ is the $N \times 1$ received vector on the $l^{\mathrm{th}}$ receive antenna. Assuming uncorrelated receive antennas, we have 
\begin{equation} \label{eq:bitwise_likelihood}
    p(\mathbf{Y}|b_i = 1) = \prod_{l=1}^{N_r} p(\mathbf{y}_l|b_i = 1) = \prod_{i=1}^{N_r} {\frac{\sum_{m \in \mathcal{C}_{b_i=1}} p(\mathbf{y}_l|\mathcal{H}_m) \pi_m}{p}}.
\end{equation}
In order to compute $p(\mathbf{y}_l|b_i = 1)$, we need to compute $p(\mathbf{y}_l|\mathcal{H}_m)$ accurately. While doing so, we can either  perform joint channel estimation and detection, which we will refer to as a noncoherent receiver design (since we do not explicitly estimate the channel) or we can first estimate the channel from the pilot REs and then perform detection as part of a coherent receiver design.

\subsection{Optimal Noncoherent Receiver} \label{subsec:optimal_noncoherent}
Suppose that $\mathbf{h}_l\in \mathcal{CN}(\mathbf{0},\mathbf{R_{hh}})$ (aligns with any TDL channel model in \cite{3gpp.38.901}) and $\mathbf{n}_l \in \mathcal{CN}(\mathbf{0}, \sigma^2 \mathbf{I}_N)$. In that case, $\mathbf{y}_l$ conditioned on $\mathcal{H}_m$ is a complex normal Gaussian random vector \cite{neeser2002proper} with
\begin{equation} \label{eq:codeword_likelihood}
    p(\mathbf{y}_l|\mathcal{H}_m) = \frac{1}{\pi^N \mathrm{det} \mathbf{\Sigma}} \mathrm{exp}(-\mathbf{y}^H_l \mathbf{\Sigma}^{-1}  \mathbf{y}_l)
\end{equation}
where $\mathbf{\Sigma}$ is the covariance of $\mathbf{y}_l$ conditioned on $\mathcal{H}_m$ and is given by
\begin{equation}
    \mathbf{\Sigma} = \mathrm{NV} (\mathrm{SNR} ~ \mathrm{diag}(\mathbf{c}_{p+d,m}) \mathbf{R_{hh}} \mathrm{diag}(\mathbf{c}_{p+d,m})^H + \mathbf{I}_N).
\end{equation}
This follows from the defintion of $\mathbf{y}_l$ in \eqref{eq:received_signal} and then computing $\mathbf{\Sigma} = \mathbb{E}[\mathbf{y}_l\mathbf{y}_l^H]$.

\subsection{Optimal Coherent Receiver} \label{subsec:optimal_coherent}
Observe that $p(\mathbf{y}_l|\mathcal{H}_m)$ can be written as $p(\mathbf{y}_{p,l}, \mathbf{y}_{d,l}|\mathcal{H}_m)$. By Bayes rule, we have
\begin{equation} \label{eq:bayes}
    p(\mathbf{y}_{p,l}, \mathbf{y}_{d,l}|\mathcal{H}_m) = p(\mathbf{y}_{d,l}|\mathcal{H}_m, \mathbf{y}_{p,l}) p(\mathbf{y}_{p,l} | \mathcal{H}_m)
\end{equation}

Denote by $\mathbf{\tilde{h}}_{d,l} = g \mathbf{h}_{d,l}$. In the equations to follow, we will utilize the subscript $1$ to denote $\mathbf{\tilde{h}}_{d,l}$ and $2$ to denote $\mathbf{y}_{p,l}$ for notational brevity. Conditioned on the $m^{th}$ hypothesis $\mathcal{H}_m$, the distribution of $\mathbf{\tilde{h}}_{d,l}$ conditional on $\mathbf{y}_{p,l} = \mathbf{y}_p$ is a multivariate Gaussian \cite{eaton2007home} such that
\begin{equation}
    \mathbf{\tilde{h}}_{d,l} = \mathbb{E}[\mathbf{\tilde{h}}_{d,l}|\mathcal{H}_m,\mathbf{y}_{p,l} = \mathbf{y}_p] + \mathrm{Cov}^{\frac{1}{2}}[\mathbf{\tilde{h}}_{d,l}|\mathcal{H}_m,\mathbf{y}_{p,l} = \mathbf{y}_p] \mathbf{z},
\end{equation}
where $\mathbf{z} \sim \mathcal{CN}(\mathbf{0},\mathbf{I}_{N_d})$ and 
\begin{equation} \label{eq:mu_sigma_1}
\begin{gathered}
    \mathbb{E}[\mathbf{\tilde{h}}_{d,l}|\mathcal{H}_m,\mathbf{y}_{p,l} = \mathbf{y}_p] = \mathbf{\Sigma}_{12} \mathbf{\Sigma}^{-1}_{22} \mathbf{y}_p \\
    \mathrm{Cov}[\mathbf{\tilde{h}}_{d,l}|\mathcal{H}_m,\mathbf{y}_{p,l} = \mathbf{y}_p] = \mathbf{\Sigma}_{11} - \mathbf{\Sigma}_{12}\mathbf{\Sigma}^{-1}_{22} \mathbf{\Sigma}_{21}.
\end{gathered} 
\end{equation}
Utilizing the definition of the received signal in \eqref{eq:received_signal}, we can compute $\mathbf{\Sigma}_{11}$, $\mathbf{\Sigma}_{12}$ and $\mathbf{\Sigma}_{22}$ (refer Appendix \ref{subsec:conditional_gaussian} for derivation) as
\begin{equation} \label{eq:Sigma}
\begin{gathered}
    \mathbf{\Sigma}_{11} = g^2 \mathbf{R}_{\mathbf{h}_d \mathbf{h}_d}\\
    \mathbf{\Sigma}_{12} = g^2 \sqrt{\beta_m} \mathbf{R}_{\mathbf{h}_d \mathbf{h}_p}\\
    \mathbf{\Sigma}_{22} = g^2 \beta_m \mathbf{R}_{\mathbf{h}_p \mathbf{h}_p} + \sigma^2 \mathbf{I}_{N_p}.
\end{gathered}
\end{equation}
From \eqref{eq:received_signal}, we observe that $\mathbf{y}_{d,l} = \sqrt{\alpha}_m\mathbf{\tilde{h}}_{d,l} \odot \mathbf{c}_m + \mathbf{n}_{d,l}$. Hence $\mathbf{y}_{d,l}$ conditioned on $\mathcal{H}_m$ and $\mathbf{y}_{p,l} = \mathbf{y}_p$ is a complex Gaussian vector with 
\begin{equation}
\begin{gathered}
     \boldsymbol{\mu}_{d,l} = \sqrt{\alpha_m} \mathbb{E}[\mathbf{\tilde{h}}_{d,l}|\mathcal{H}_m,\mathbf{y}_{p,l} = \mathbf{y}_p] \odot \mathbf{c}_m\\
    \mathbf{\Sigma}_{d,l} = \alpha_m \mathrm{diag}(\mathbf{c}_m) \mathrm{Cov}[\mathbf{\tilde{h}}_{d,l}|\mathcal{H}_m,\mathbf{y}_{p,l} = \mathbf{y}_p] \mathrm{diag}(\mathbf{c}_m)^H + \sigma^2 \mathbf{I}_{N_d},
\end{gathered} 
\end{equation}
with the dependence of $\boldsymbol{\mu}_{d,l}$ and $\mathbf{\Sigma}_{d,l}$ on $\mathcal{H}_m$ being kept implicit notationally. By substituting for $\mathbf{\Sigma}_{11}$, $\mathbf{\Sigma}_{12}$ and $\mathbf{\Sigma}_{22}$  in \eqref{eq:mu_sigma_1} and factoring out $\mathrm{SNR}$, we obtain
\begin{equation} \label{eq:mu_sigma}
\begin{gathered}
    \boldsymbol{\mu}_{d,l} = \sqrt{\frac{\alpha_m}{\beta_m}}  \mathbf{W}_m \mathbf{y}_p \odot \mathbf{c}_m \\
    \mathbf{\Sigma}_{d,l} =  \mathrm{NV} \big(\mathbf{I}_{N_d} + \alpha_m\mathrm{SNR} ~ \mathrm{diag}(\mathbf{c}_m)\big(\mathbf{R}_{\mathbf{h}_d \mathbf{h}_d} - \mathbf{W}_m \mathbf{R}_{\mathbf{h}_p \mathbf{h}_d} \big) \mathrm{diag}(\mathbf{c}_m)^H \big),
\end{gathered}
\end{equation}
where 
\begin{equation} \label{eq:chest}
    \mathbf{W}_m = \mathbf{R}_{\mathbf{h}_d \mathbf{h}_p} \bigg(\mathbf{R}_{\mathbf{h}_p \mathbf{h}_p} + \frac{1}{\beta_m \mathrm{SNR}} \mathbf{I}_{N_p}\bigg)^{-1}.
\end{equation}
Observe that by plugging in $\beta_m = 1$ or $\beta_m = \alpha_m$ in \eqref{eq:mu_sigma}, $\mathbf{W}_m \mathbf{y}_p$ is the linear minimum mean squared error (LMMSE) channel estimate with the $\mathrm{SNR}$ appropriately shaped by $\beta_m$. Having computed $\boldsymbol{\mu}_{d,l}$ and $\mathbf{\Sigma}_{d,l}$ in \eqref{eq:mu_sigma}, we can now compute the first term in the RHS of \eqref{eq:bayes} as
\begin{equation} \label{eq:coherent_prob}
    p(\mathbf{y}_{d,l}|\mathcal{H}_m, \mathbf{y}_{p,l}) = \frac{1}{\pi^{N_d} ~\mathrm{det}\mathbf{\Sigma}_{d,l}} \exp{\big(-(\mathbf{y}_{d,l} - \boldsymbol{\mu}_{d,l})^H \mathbf{\Sigma}^{-1}_{d,l} (\mathbf{y}_{d,l} - \boldsymbol{\mu}_{d,l})\big)}.
\end{equation}
From \eqref{eq:received_signal}, we observe that $\mathbf{y}_{p,l} = \sqrt{\beta_m}\mathbf{\tilde{h}}_{p,l} \odot \mathbf{1} + \mathbf{n}_{p,l}$. Conditioned on $\mathcal{H}_m$, $\mathbf{y}_{p,l}$ is a complex Gaussian vector with
\begin{equation}
\begin{gathered}
    \boldsymbol{\mu}_{p,l} = \mathbf{0} \\
    \mathbf{\Sigma}_{p} = \mathrm{NV}(\mathbf{I}_{N_p} + \beta_m\mathrm{SNR} ~\mathbf{R}_{\mathbf{h}_p \mathbf{h}_p}),
\end{gathered}
\end{equation}
and hence we can compute the second term in the RHS of \eqref{eq:bayes} as
\begin{equation}
    p(\mathbf{y}_{p,l} | \mathcal{H}_m) = \frac{1}{\pi^{N_p} ~\mathrm{det} \mathbf{\Sigma}_p} \exp{\big(-\mathbf{y}^H_{p,l} \mathbf{\Sigma}_p^{-1} \mathbf{y}^H_{p,l}\big)}.
\end{equation}
Observe that if $\beta_m = 1$ or any other constant independent of $\mathcal{H}_m$, then $p(\mathbf{y}_{p,l} | \mathcal{H}_m) = p(\mathbf{y}_{p,l})$. In current NR standards \cite{3gpp.38.212}, the DMRS is always transmitted at constant power. Under that assumption, we have 
\begin{equation}
    \arg \max_{m} p(\mathbf{y}_l|\mathcal{H}_m) = \arg \max_{m} p(\mathbf{y}_{d,l}|\mathcal{H}_m, \mathbf{y}_{p,l}).
\end{equation}
Note that $p(\mathbf{y}_{d,l}|\mathcal{H}_m, \mathbf{y}_{p,l})$ is computed by first computing the LMMSE channel estimate $\hat{\mathbf{h}}_l = \sqrt{\alpha_m / \beta_m} \mathbf{W}_m \mathbf{y}_p$ and then substituting it in \eqref{eq:coherent_prob}. This motivates the rationale behind coherent communication in practice -- we perform channel estimation and then utilize the channel estimate on the data REs to decode and demodulate the data REs\footnote{Note that \eqref{eq:coherent_prob} computes $\Sigma_{d,l}$ optimally as well which might not be done in practice.}. However, if the DMRS is subject to data dependent power shaping, then $p(\mathbf{y}_{p,l} | \mathcal{H}_m)$ is no longer independent of $\mathcal{H}_m$ and simply computing $p(\mathbf{y}_{d,l}|\mathcal{H}_m, \mathbf{y}_{p,l})$ is no longer optimal. 

\subsection{Low Complexity Coherent Receiver} \label{subsec:lc_approx}
Both the receivers proposed in Section \ref{subsec:optimal_noncoherent} and \ref{subsec:optimal_coherent} compute identical probabilities. However, the computation of the bitwise likelihood in \eqref{eq:bitwise_likelihood} using the receivers described thus far will require computing a weighted sum of $2^{k-1}$ exponentials for each of the $k$ information bits. In order to reduce the computational complexity, we demonstrate how to simplify the coherent receiver design described in Section \ref{subsec:optimal_coherent} assuming $\beta_m = 1$ and $\mathbf{R}_{\mathbf{hh}} = \mathbf{1}_N\mathbf{1}_N^T$ i.e., channel is frequency flat. Our results in Section \ref{subsec:results_prac_coh_noncoh} will demonstrate that we can drop $p(\mathbf{y}_{p,l}|\mathcal{H}_m)$ even if $\beta_m \neq 1$ without significant performance degradation. However, the second assumption will cause minimal performance degradation only if the number of resource blocks (RBs) allocated is small as is the case for PUCCH, so that the channel is effectively ``frequency-flat''.

Substituting for $p(\mathbf{y}_{p,l}, \mathbf{y}_{d,l}|\mathcal{H}_m)$ in $p(\mathbf{y}_l | b_i = 1)$, we have
\begin{equation}
    p(\mathbf{y}_l | b_i = 1) \propto \frac{\sum_{m \in \mathcal{C}_{b_i=1}} \frac{\pi_m}{\mathrm{det} \mathbf{\Sigma}_{d,l}} \exp{\big(-(\mathbf{y}_{d,l} - \boldsymbol{\mu}_{d,l})^H \mathbf{\Sigma}^{-1}_{d,l} (\mathbf{y}_{d,l} - \boldsymbol{\mu}_{d,l})\big)}}{p}.
\end{equation}
Approximating the sum of exponentials by the dominating exponential and taking $\log$ on both sides, we have
\begin{equation} \label{eq:bit_likelihood}
    \log{p(\mathbf{y}_l , b_i = 1)} \propto \max_{m \in \mathcal{C}_{b_i=1}} -\Delta^2(\mathbf{y}_{d,l},\boldsymbol{\mu}_{d,l}) + \log{\frac{\pi_m}{\mathrm{det} \mathbf{\Sigma}_{d,l}}},
\end{equation}
where
\begin{equation} \label{eq:mlb_dist}
    \Delta^2(\mathbf{y}_{d,l},\boldsymbol{\mu}_{d,l}) = (\mathbf{y}_{d,l} - \boldsymbol{\mu}_{d,l})^H \mathbf{\Sigma}^{-1}_{d,l} (\mathbf{y}_{d,l} - \boldsymbol{\mu}_{d,l})
\end{equation}
is the Mahalanobis distance between $\mathbf{y}_{d,l}$ and $\boldsymbol{\mu}_{d,l}$. By substituting for flat fading in $\mathbf{R}_{\mathbf{hh}}$ and repeatedly utilizing the matrix inversion lemma \eqref{eq:inv_lemma_simple} (refer Appendix \ref{subsec:lc_inv_lemma} for the proof), we obtain
\begin{equation} \label{eq:lc_inv_lemma}
\begin{gathered}
    \mathrm{det} \mathbf{\Sigma}_{d,l} = \mathrm{NV}^{N_d}\bigg(1 + \frac{\alpha_mN_d}{\beta_mN_p + \frac{1}{\mathrm{SNR}}} \bigg) \\
    \Delta^2(\mathbf{y}_{d,l},\boldsymbol{\mu}_{d,l}) = \frac{1}{\mathrm{NV}} \bigg(\|\mathbf{y}_{d,l} - \boldsymbol{\mu}_{d,l}\|^2_2 - \frac{\alpha_m}{\alpha_m N_d + \beta_m N_p + \frac{1}{\mathrm{SNR}}} \|(\mathbf{y}_{d,l} - \boldsymbol{\mu}_{d,l})^H \mathbf{c}_m \|^2_2 \bigg).
\end{gathered}
\end{equation}
Observe that the definition of $\Delta^2(\mathbf{y}_{d,l},\boldsymbol{\mu}_{d,l})$ in \eqref{eq:lc_inv_lemma} contains both $2 \Re(\mathbf{y}_{d,l}^H \boldsymbol{\mu}_{d,l})$ (from $\|\mathbf{y}_{d,l} - \boldsymbol{\mu}_{d,l}\|^2_2$)  and $\|\mathbf{y}_{d,l}^H \mathbf{c}_m\|^2_2$ (from $\|(\mathbf{y}_{d,l} - \boldsymbol{\mu}_{d,l})^H \mathbf{c}_m \|^2_2$). Recalling that $\boldsymbol{\mu}_{d,l}$ is an elementwise product of the channel estimate and the codeword $\mathbf{c}_m$, the first term models traditional coherent communication e.g. codeword ML decoder. Meanwhile, the second term computes the energy of $\mathbf{y}_{d,l}$ when projected onto $\mathbf{c}_m$, which is typically computed by an energy detector used for non-coherent communication.

Having computed $\mathrm{det} \mathbf{\Sigma}_{d,l}$ and $\Delta^2(\mathbf{y}_{d,l},\boldsymbol{\mu}_{d,l})$, we can substitute them back in \eqref{eq:bit_likelihood}, compute $\log{p(\mathbf{Y}|b_i = j)}$ for $j \in \{0,1\}$ using \eqref{eq:bitwise_likelihood} and utilize the Bitwise UEP rule in \eqref{eq:rule_beta} to decode $\hat{b}_i = 0$ if 
\begin{equation} \label{eq:lc_bitwise_rule}
    \log{p(\mathbf{Y} | b_i = 1)} - \log{p(\mathbf{Y} | b_i = 0)} \geq \log{\frac{\beta}{(1 - \beta)}}.
\end{equation}
\vspace{2mm}
\textbf{Computational Complexity Analysis}: Returning to \eqref{eq:bit_likelihood}, we observe that $\log{\frac{\pi_m}{\mathrm{det} \mathbf{\Sigma}_{d,l}}}$ can be precomputed $\forall ~ m \in [2^k]$. More precisely, $\log{\frac{\pi_m}{\mathrm{det} \mathbf{\Sigma}_{d,l}}} + \log{\mathrm{NV}}$ is independent of the received signal (since we will fix SNR used in receiver processing to a nominal value independent of the received SNR as will be detailed in Section \ref{subsec:snr_ne_comp}). Given $\mathbf{y}_{d,l}$, $\boldsymbol{\mu}_{d,l}$ and $\mathbf{c}_m$ are vectors of length $n/2$, computing $\Delta^2(\mathbf{y}_{d,l},\boldsymbol{\mu}_{d,l})$ requires $9n/2 + 2$ MAC operations. Furthermore, we compute $\Delta^2(\mathbf{y}_{d,l},\boldsymbol{\mu}_{d,l})$ for each of the $N_r$ receive antennas and $2^k$ codeword hypothesis -- $2^kN_r(9n/2 + 2)$. Finally, to evaluate \eqref{eq:lc_bitwise_rule}, we have to compute a max over the $2^{k-1}$ values of $ -\Delta^2(\mathbf{y}_{d,l},\boldsymbol{\mu}_{d,l}) + \log{\frac{\pi_m}{\mathrm{det} \mathbf{\Sigma}_{d,l}}}$ for $b_i = 0$ and $b_i =1$, and repeat this for each of the $k$ information bits -- $k2^k$ operations. Hence, the computational complexity is $2^k (N_r(9n/2 + 2) + k))$ MAC operations. An ML decoder computes $\sum_{l} \mathbf{y}^H_{d,l} (\mathbf{\hat{h}}_{d,l} \odot \mathbf{c}_m)$ for each codeword $\mathbf{c}_m$, which requires $2^k(2nN_r + 1)$ MAC operations. Hence for $k \ll nN_r$, the legacy ML decoder and the low complexity coherent receiver have comparable computational complexities.

\section{Simulation Details} \label{sec:simulation_details}
In this section, we will cover the simulation details that are common to both the AWGN receiver design (that assumes perfect channel estimation) and the fading channel receiver design. In the subsequent sections, we will first highlight the simulation details specific to each, before diving into the results.

\subsection{Probabilistic Model for HARQ-ACK distribution} \label{subsec:prob_model}
\begin{figure}
    \centering
    \includegraphics[width = 4in]{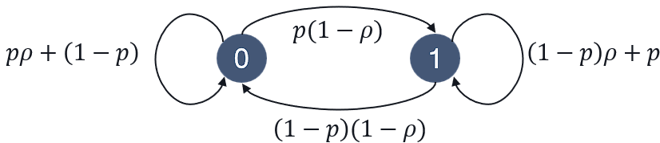}
    \caption{Transition probabilities for consecutive HARQ-ACK bits}
    \label{fig:markov_chain}
\end{figure}

As mentioned in Section \ref{subsec:mlm_glrt}, owing to downlink channel statistics, HARQ-ACK bits in a single PUCCH transmission can be correlated. To model the correlation among bits within a single PUCCH transmission, we consider a first order Markov model, wherein the ACK probability $p(b_i = 1) = p$ and $\rho \in [0,1]$ is the correlation coefficient between consecutive bits $b_{i-1}$ and $b_i$. The transition probabilities are given by
\begin{equation} \label{eq:trans_prob}
\begin{aligned}
    p(b_i = 1|b_{i-1} = 0) &= p(1 - \rho) \\
    p(b_i = 0|b_{i-1} = 1) &= (1 - p)(1 - \rho), \\
\end{aligned}
\end{equation}
and the transition diagram is illustrated in Fig. \ref{fig:markov_chain}. Observe that setting $\rho = 0$ makes the bits i.i.d (since $p(b_i = 1|b_{i-1} = 0) = p(b_i = 1)$). On the other hand, setting $\rho = 1$ implies only two sequences are seen - all-ACK with probability $p$ and all-NACK with probability $1 - p$. More generally, given $p$ and $\rho$, the codeword probability $\pi_m$ corresponding to the message sequence $\mathbf{b}_m = \{b_0, b_1 \ldots b_{k-1}\}$ is given by
\begin{equation}
    \pi_m = p(b_0) \Pi_{i=1}^{k-1} p(b_i|b_{i-1}).
\end{equation}
While it can be argued that a first order Markov model may not be sufficient to capture the correlation among HARQ-ACK bits, PDSCH CRC logs from field measurements have supported the adequacy of first order Markov modeling. 

\subsection{5G NR implementation}
Coding for short block lengths is described in \cite{3gpp.38.212}. Short block lengths refers to the $3 \leq k \leq 11$ regime, where $k$ is the number of information bits. For this regime, NR provides a generator matrix $\mathbf{G} \in \{0,1\}^{32 \times 11}$ (refer Table 5.3.3.3-1 in \cite{3gpp.38.212}) which when multiplied by the message sequence $\mathbf{b}_m$ as $\mathbf{Gb}_m ~ \mathrm{mod} ~ 2$ yields the coded sequence. Given the rate tuple $(k,n)$ for $n \leq 32$, we select the first $k$ columns and first $n$ rows of $\mathbf{G}$ as the generator matrix. ML decoding is employed in current standard evaluations to compute the BLER for short block length NR codes.

PUCCH HARQ-ACK is QPSK modulated prior to transmission in NR (refer Section 6.3.2.5.2 in \cite{3gpp.38.211}) in order to achieve the target BLER at low SNRs. Considering an OFDM waveform, this is equivalent to mapping two coded bits to a QPSK constellation point before placing it on a resource element (RE) in an OFDM grid. By analogy, if we do not quantize the output of the transformer to the binary field (GF(2)), we can view the proposed encoder design as a JSCCM (M for Modulation) scheme wherein two coded symbols (two real values) are mapped to a single complex number and placed on one RE. Since the coded symbols are real valued, we cannot define a code rate but we ensure that the number of information bits per dimension (i.e. per RE), and hence the spectral efficiency, are the same for the baseline and proposed schemes.

\subsection{UEP Criterion} \label{subsec:uep_criterion}
According to Section 8.3 in \cite{3gpp.38.104}, the ACK and NACK error rate of PUCCH decoding should not exceed $1\%$ and $0.1\%$ respectively. Denote by $\mathrm{SNR}_{\mathrm{ack}}$ the minimum SNR required to achieve an ACK error rate of $\leq 1\%$ and by $\mathrm{SNR}_{\mathrm{nack}}$ the minimum SNR required to achieve a NACK error rate of $\leq 0.1\%$. Compute $\mathrm{SNR}_{\mathrm{th}} = \max \{\mathrm{SNR}_{\mathrm{ack}}, \mathrm{SNR}_{\mathrm{nack}} \}$ (taking the maximum ensures both ACK and NACK error rate targets are met). $\mathrm{SNR}_{\mathrm{th}}$ will serve as the performance metric to compare the different encoder and decoder designs.

\subsection{Loss Functions \& Training Algorithms} \label{subsec:training_algo}
As part of the transformer training, the transformer will encode a batch of message sequences which are subsequently power shaped, passed through a SISO AWGN channel and decoded assuming perfect channel estimation. Since we are focused on the unequal protection of ACK and NACK bit errors as introduced in Section \ref{sec:awgn_receiver}, our training objective should attempt to lower both the ACK and NACK bit error rate. Let us first focus on reducing $\mathrm{BER}_{\mathrm{ACK}}$ at a given SNR. Observe that
\begin{equation}
\begin{aligned}
    1 - \mathrm{BER}_{\mathrm{ACK}} &= \int 1_{\mathbf{y} \in \Gamma_1} p(\mathbf{y}|b_i = 1) d\mathbf{y} \\
    &= \int 1_{\mathbf{y} \in \Gamma_1} \frac{p(b_i = 1|\mathbf{y})}{p} p(\mathbf{y}) d\mathbf{y} \\
    &= \mathbb{E}_{\mathbf{y}} \bigg[ 1_{\mathbf{y} \in \Gamma_1} \frac{p(b_i = 1|\mathbf{y})}{p}\bigg],
\end{aligned}
\end{equation}
where $\Gamma_1$ denotes the decision region where $b_i$ is decoded to be 1, i.e.
\begin{equation}
\Gamma_1 = \left\{ {\bf y} ;  \frac{p({\bf y}| b_i=0)}{p({\bf y} | b_i=1)} < \frac{\beta}{1-\beta}\right\},
\end{equation}
and the posterior bit probability is computed as described in \eqref{eq:bitwise_MAP}. 

Hence we want to maximize $1_{\mathbf{y} \in \Gamma_1} p(b_i = 1|\mathbf{y})$, where both $\Gamma_1$ and $p(b_i = 1|\mathbf{y})$ are a function of the encoder weights $\boldsymbol{\theta}$. Given that $1_{\mathbf{y} \in \Gamma_1}$ would have zero gradient\footnote{Subgradient at points of discontinuity would also be 0.} w.r.t $\boldsymbol{\theta}$ for all $\mathbf{y}$, it is simpler to maximize $1_{b_i =1} p(b_i = 1|\mathbf{y})$. Note that bit $b_i$ is decoded correctly w.h.p above a certain SNR, implying that for most $\mathbf{y}$, $1_{b_i =1} = 1_{\mathbf{y} \in \Gamma_1}$. Given a training SNR aimed at achieving BER of $\leq 1\%$, the approximation holds true while training the encoder. Similarly, for reducing $\mathrm{BER}_{\mathrm{NACK}}$, we maximize $1_{b_i =0} p(b_i = 0|\mathbf{y})$. Hence, we train the encoder weights $\boldsymbol{\theta}$ to maximize
\begin{equation} \label{eq:loss_fn}
    \mathcal{L}(\mathbf{y},\{t_i\};\boldsymbol{\theta}) = \mathbb{E}_{\mathbf{y}} \bigg[ \sum_{i=0}^{k-1} (1 - t_i) p_{\boldsymbol{\theta}}(b_i = 0|\mathbf{y}) + t_i p_{\boldsymbol{\theta}}(b_i = 1|\mathbf{y}) \bigg]
\end{equation}
where $t_i = 1$ if $b_i = 1$ else it is 0. Observe that both the bitwise posterior probabilities are equally weighted in \eqref{eq:loss_fn}, since we did not observe any performance improvement by unequal weighting of the posterior probabilities in simulation.

We consider two approaches to training the transformer-based codebook - Scenario-Specific (SS) and Free-Lunch (FL). SS training refers to learning a different codebook for every distribution, while FL training refers to the learning of a common codebook for $\{\mathbf{\Pi}_0, \mathbf{\Pi}_1, \ldots \mathbf{\Pi}_{N-1}\}$. The FL training is outlined in Algorithm \ref{alg:free_lunch_training}. It reduces to SS training by setting $N=1$. We do not learn the power shaping as part of the transformer training. Instead we fix the power shaping scheme to one of the three techniques presented in Section \ref{sec:ps}, and learn the optimal normalized codebook $\mathcal{C}$ given the PS scheme (learning PS as part of transformer training is covered in Appendix \ref{subsec:learnt_ps_appendix}). If the common codebook were to be learnt for a very diverse set of distributions e.g. $p = 0.5$ and $p = 0.9$, then a scenario-specific codebook would outperform the FL codebook as expected. However, if one were to train the common codebook on a set of closely-related distributions, then the performance can improve marginally as will be demonstrated in Section \ref{subsec:ss_fl_results}. Hence we refer to the latter technique as ``free-lunch" -- it not only provides for learning of a robust codebook but also has the potential to yield improved performance. 

\vspace{2mm}
\begin{remark} \label{remark:free_lunch}
\textit{The idea of free-lunch training draws from domain randomization in robotics \cite{tobin2017domain}, a simple technique for training models on simulated images that transfer to real images by randomizing rendering in the simulator. Training on a set of closely-related distributions can be thought of as being analogous to domain randomization.}
\end{remark}
\vspace{2mm}

\begin{algorithm} 
\DontPrintSemicolon
\SetAlgoHangIndent{0pt}
\setstretch{1}
\For{number of training iterations}{
\For{$\mathbf{\Pi} \in \{\mathbf{\Pi}_0, \mathbf{\Pi}_1, \ldots \mathbf{\Pi}_{N-1}\}$}{
Compute $\mathrm{SNR}(\mathbf{\Pi})$ using \eqref{eq:snr_entropy}. Set $g^2/\sigma^2 = \mathrm{SNR}(\mathbf{\Pi})$.\;
Sample minibatch of $N_{\mathrm{batch}}$ messages $\mathbf{B} = \{\mathbf{b}_m\} \sim \mathbf{\Pi}$.\;
\For{$\mathbf{b}_m \in \mathbf{B}$}{
Set target labels $t_i = b_i$.\;
Encode (and modulate) $\mathbf{b}_m$ using the transformer to obtain $\mathbf{c}_{m}$.\;
Apply power shaping to obtain $\sqrt{\alpha_{m}}\mathbf{c}_{m}$.\;
Compute $\mathbf{y} = g\sqrt{\alpha_{m}}\mathbf{c}_{m} + \mathbf{n}$ where $\mathbf{n} \sim \mathcal{CN}(0,\mathbf{I}_{n/2})$.\;
\For{$i \in [k]$}{
Compute $p_{\mathbf{\theta}}(b_i=1|\mathbf{y})$ using \eqref{eq:bitwise_MAP}.
}}
Compute $\mathcal{L}(\mathbf{y},\{t_i\};\boldsymbol{\theta})$ in \eqref{eq:loss_fn} using $p_{\boldsymbol{\theta}}(b_i=1|\mathbf{y})$ and $t_i$. \\
$\boldsymbol{\theta} = \boldsymbol{\theta} - \gamma \mathrm{AdamW}(\nabla_{\theta} \mathcal{L}(\mathbf{y},\{t_i\};\boldsymbol{\theta}))$
}
}
\caption[caption]{Free Lunch Training} \label{alg:free_lunch_training}
\end{algorithm}

For both techniques, we validate on the ACK and NACK BER at $p = 0.9, \rho =0$. Furthermore, it is essential to train the transformer at an $\mathrm{SNR}$ where the ACK and NACK BERs are approximately $0.1\%$ and $1\%$ respectively (since the decoder computes $p(b_i = 0|\mathbf{y})$, this is equivalent to bitwise MAP, hence ACK is protected more than NACK). In order to adjust the training SNR for different distributions, we compute $\mathrm{SNR}_{\mathrm{ref}}$ for one distribution $\mathbf{\Pi}_{\mathrm{ref}}$ by hand-tuning. Given distribution $\mathbf{\Pi}$, we set $\mathrm{SNR}(\mathbf{\Pi})$ to be
\begin{equation} \label{eq:snr_entropy}
    \mathrm{SNR}(\mathbf{\Pi}) = \mathrm{SNR}_{\mathrm{ref}} + H(\mathbf{\Pi}) - H(\mathbf{\Pi}_{\mathrm{ref}}),
\end{equation}
where $H(\mathbf{\Pi})$ is the entropy of $\mathbf{\Pi}$
\begin{equation}
H(\mathbf{\Pi}) = - \sum_{m \in [2^k]} \pi_m \log \pi_m.
\end{equation}

\section{AWGN Receiver Results} \label{sec:awgn_receiver_results}
In this section, we set $(k,n) = (11,32)$, $A=4$ tokens and $\mathrm{SNR}_{\mathrm{ref}} = -1$ dB corresponding to the reference distribution with $p=0.9,\rho=0$ for all evaluations. 

\subsection{Computing threshold SNR $\mathrm{SNR}_{\mathrm{th}}$}
\begin{figure}
\centering
    \subfloat[ACK BER vs. SNR. $\mathrm{BER}_{\mathrm{ACK,t}} = 0.01$]{\includegraphics[width = 3in]{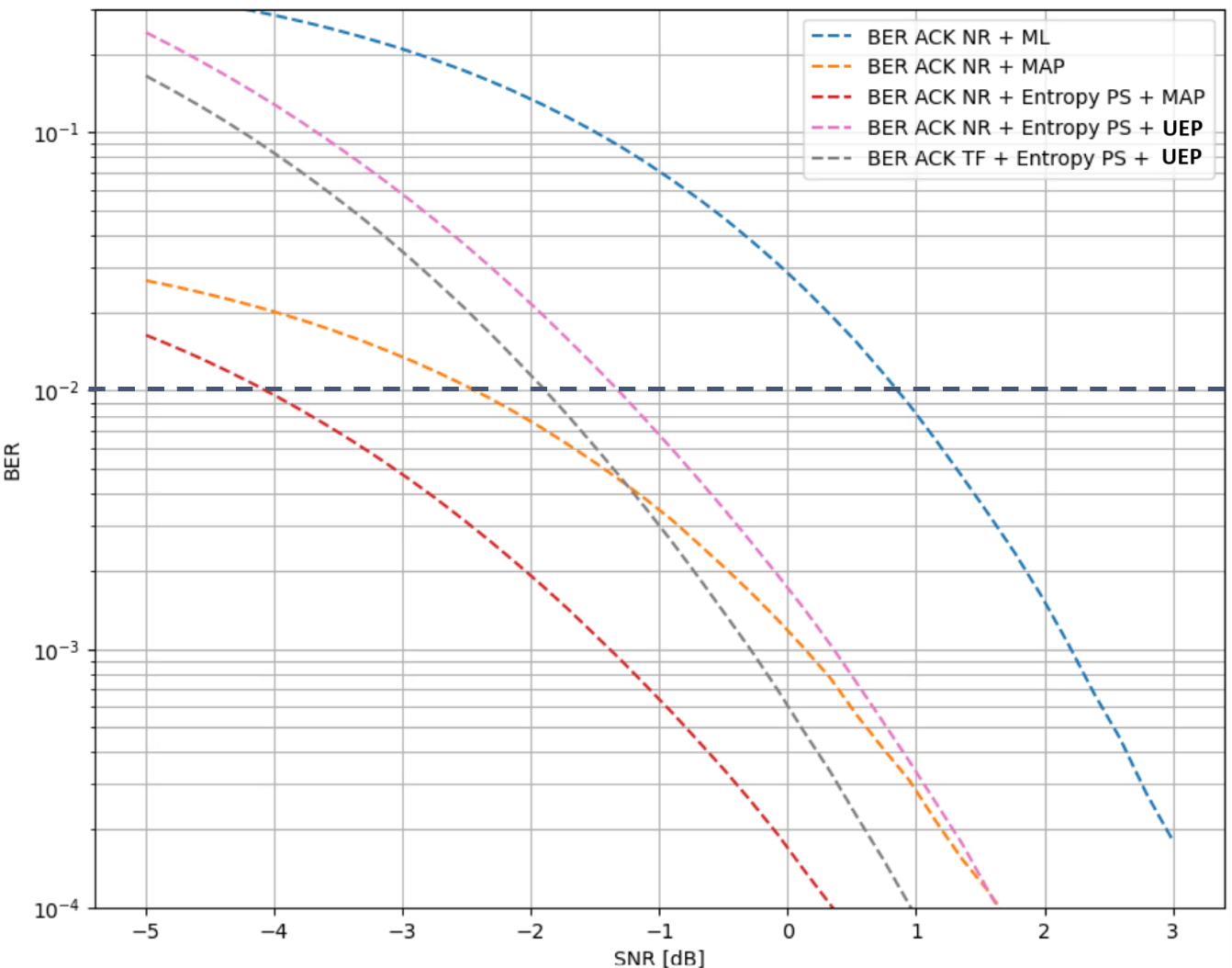}
    \label{fig:ack_ber}}
    \hspace{0.2in}
    \subfloat[NACK BER vs. SNR. $\mathrm{BER}_{\mathrm{NACK,t}} = 0.001$]{\includegraphics[width=3in]{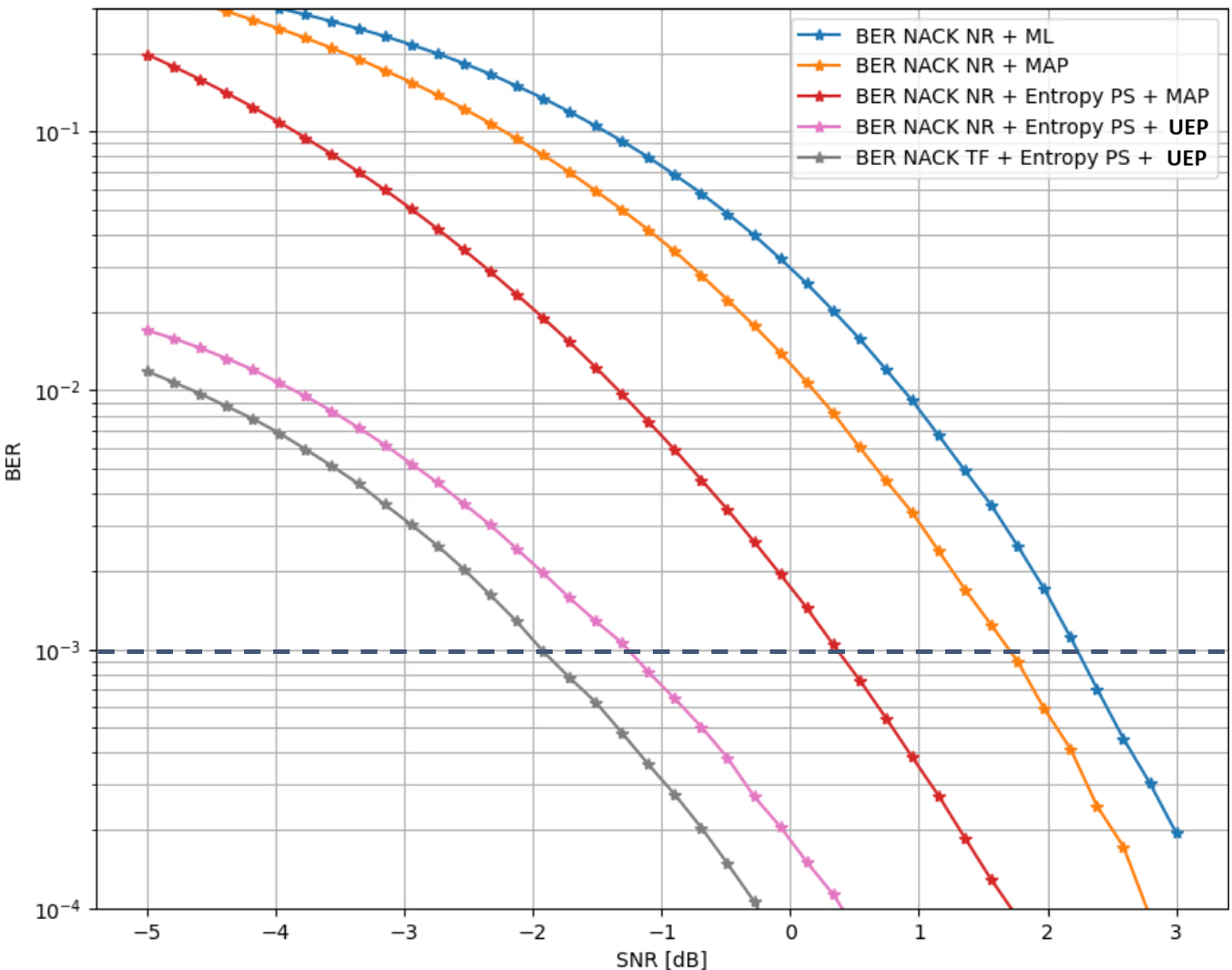}
    \label{fig:nack_ber}}
\caption{ACK and NACK BER vs. SNR for $p=0.9, \rho = 0$.}
\label{fig:ack_nack_ber}
\end{figure}

There are three ``key ingredients" in the techniques proposed thus far - the transformer-based encoder, power shaping and Bitwise UEP. To visualize the gains from each ``ingredient", we will first fix the distribution to $p = 0.9$ and $\rho = 0$ and plot the ACK and NACK BER vs. SNR in Fig. \ref{fig:ack_ber} and \ref{fig:nack_ber} respectively. To compute $\mathrm{SNR}_{\mathrm{th}}$, recall that we compute the maximum of the SNRs where $\mathrm{BER}_{\mathrm{NACK}} = 0.1\%$, where $\mathrm{BER}_{\mathrm{ACK}} = 1\%$. Observe that the NR baseline with ML decoder curve, being prior independent, is the same for ACK and NACK with the NACK curve determining $\mathrm{SNR}_{\mathrm{th}}$ to be 2.2 dB. Using a bitwise MAP decoder in place of an ML decoder reduces the ACK and NACK error rate. However, the MAP decoder preferentially protects the bit with the higher prior - ACK, hence NACK determines  $\mathrm{SNR}_{\mathrm{th}}$ to be 1.62 dB. Adding Entropy PS further drives down $\mathrm{SNR}_{\mathrm{th}}$ to 0.36 dB, thus providing a power shaping gain of 1.26 dB. Replacing the MAP decoder by a UEP decoder (approximately) swaps the ACK and NACK error curves. This can be attributed to $p/(1-p) = 9$ in the MAP decoder while $\beta/(1-\beta) = 1/10$ in the UEP decoder. This swap drives down $\mathrm{SNR}_{\mathrm{th}}$ to -1.26 dB, providing a Bitwise UEP gain of 1.62 dB. Finally, using a transformer-based encoder in place of NR reduces $\mathrm{SNR}_{\mathrm{th}}$ to -1.93 dB, providing a transformer gain of 0.67 dB.

\subsection{Effect of varying bit correlation $\rho$ on $\mathrm{SNR}_{\mathrm{th}}$}
\begin{figure}
\centering
    \subfloat[Comparing SS training with other baselines]{\includegraphics[width = 3.1in]{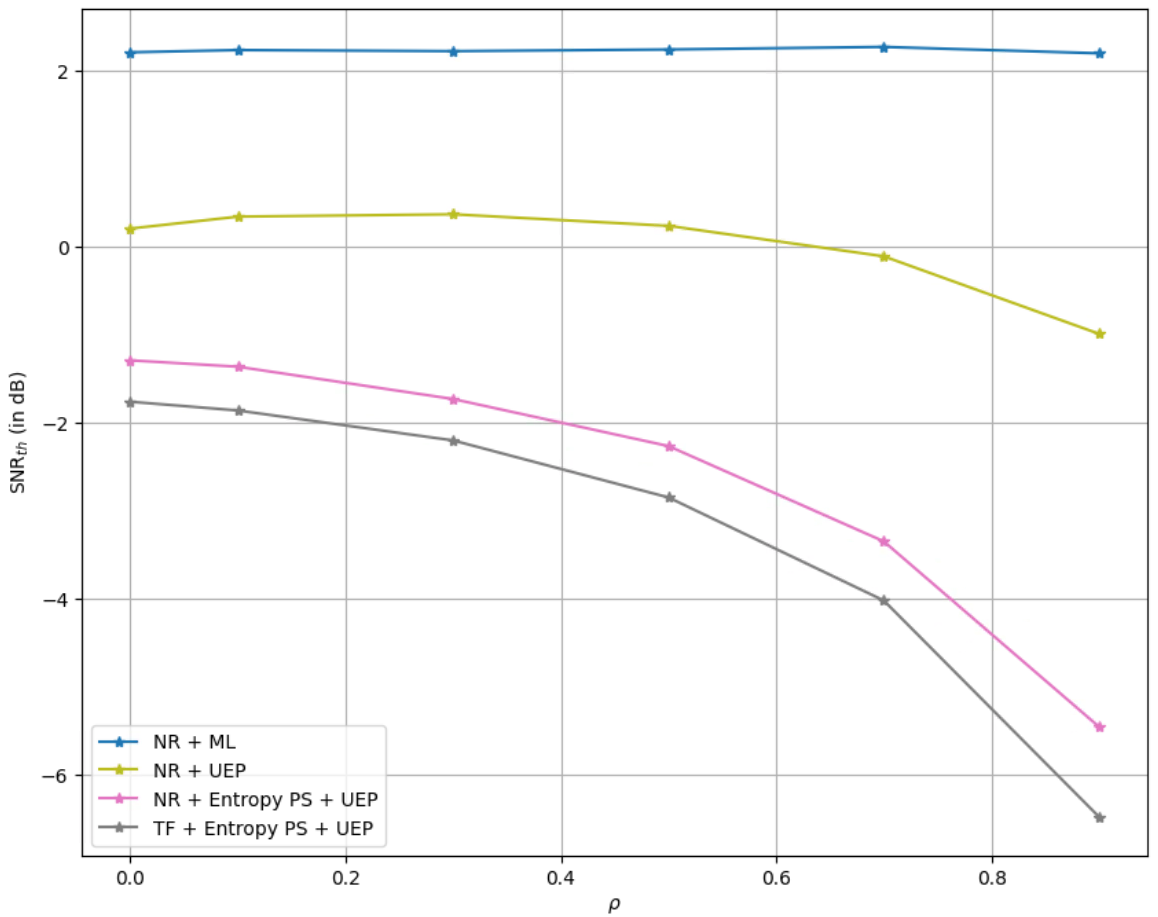}
    \label{fig:snr_th}}
    \hspace{0.01in}
    \subfloat[Comparing SS and FL training]{\includegraphics[width=3.1in]{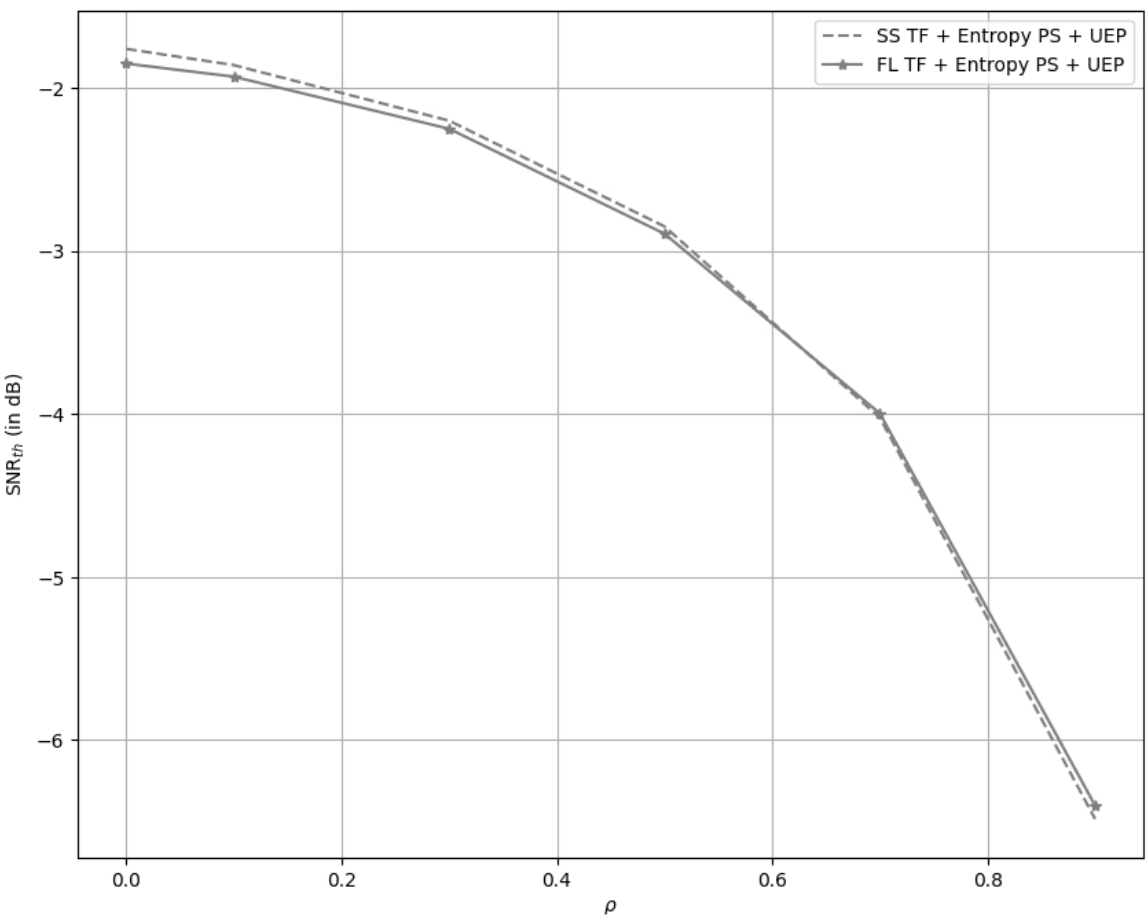}
    \label{fig:ss_fl}}
\caption{$\mathrm{SNR}_{\mathrm{th}}$ as a function of $\rho$.}
\label{fig:plots}
\end{figure}

So far, we demonstrated reduction in $\mathrm{SNR}_{\mathrm{th}}$ for one prior - $p = 0.9, \rho = 0$. However, such gains can be obtained for other priors as well. We perform Scenario-Specific (SS) training of the transformer for each distribution assuming Entropy PS, and plot $\mathrm{SNR}_{\mathrm{th}}$ vs. $\rho$ in Fig. \ref{fig:snr_th}. Since the ML decoder is prior independent, $\mathrm{SNR}_{\mathrm{th}}$ is nearly constant as a function of $\rho$ for the NR + ML baseline. Replacing the ML decoder with a UEP decoder provides a 2 dB reduction in $\mathrm{SNR}_{\mathrm{th}}$ at $\rho = 0$ and up to 3.2 dB reduction at $\rho = 0.9$. We then add Entropy PS to the NR encoder, while continuing to employ the UEP decoder. This provides a 3.5 dB reduction in $\mathrm{SNR}_{\mathrm{th}}$ at $\rho = 0$ and up to 7.6 dB reduction at $\rho = 0.9$. Replacing NR with a transformer-based encoder provides a 4.0 dB reduction in $\mathrm{SNR}_{\mathrm{th}}$ at $\rho = 0$ and up to 8.6 dB reduction at $\rho = 0.9$.

\subsection{Comparing Scenario Specific (SS) vs Free Lunch (FL) training} \label{subsec:ss_fl_results}
As discussed in Section \ref{subsec:training_algo}, FL training involves learning a common codebook for a range of closely-related distributions. We train a common codebook for $p = \{0.8, 0.85, 0.9, 0.95\}$ and $\rho = \{0, 0.1, 0.3, 0.5, 0.7, 0.9\}$. Such training also accounts for small fluctuations in the target PDSCH BLER due to fast fading or change in location of UE in the cell. For inference, we fix $p = 0.9$ and vary $\rho$. The performance of SS and FL is compared in Fig. \ref{fig:ss_fl}. FL yields small improvements of 0.1 dB at $\rho = 0$ to 0.05 dB at $\rho = 0.5$. At $\rho \geq 0.7$, there are small degradations in performance of up to $0.08$ dB. Hence, FL training enables us to learn a common codebook with negligible loss of performance.

\vspace{2mm}
\begin{remark} \label{remark:power_norm}
\textit{FL training refers only to the learning of the (normalized) codebook. The PS scheme is determined independently. For instance, with Entropy PS, the per-codeword power levels are functions of the prior, hence can be viewed as scenario-specific. However, with Arithmetic and Step PS, the power levels are prior independent and could be viewed as free-lunch.}
\end{remark}

\subsection{Comparing different PS schemes} \label{subsec:ps_results}
\begin{figure}
    \centering
    \includegraphics[width = 3.1in]{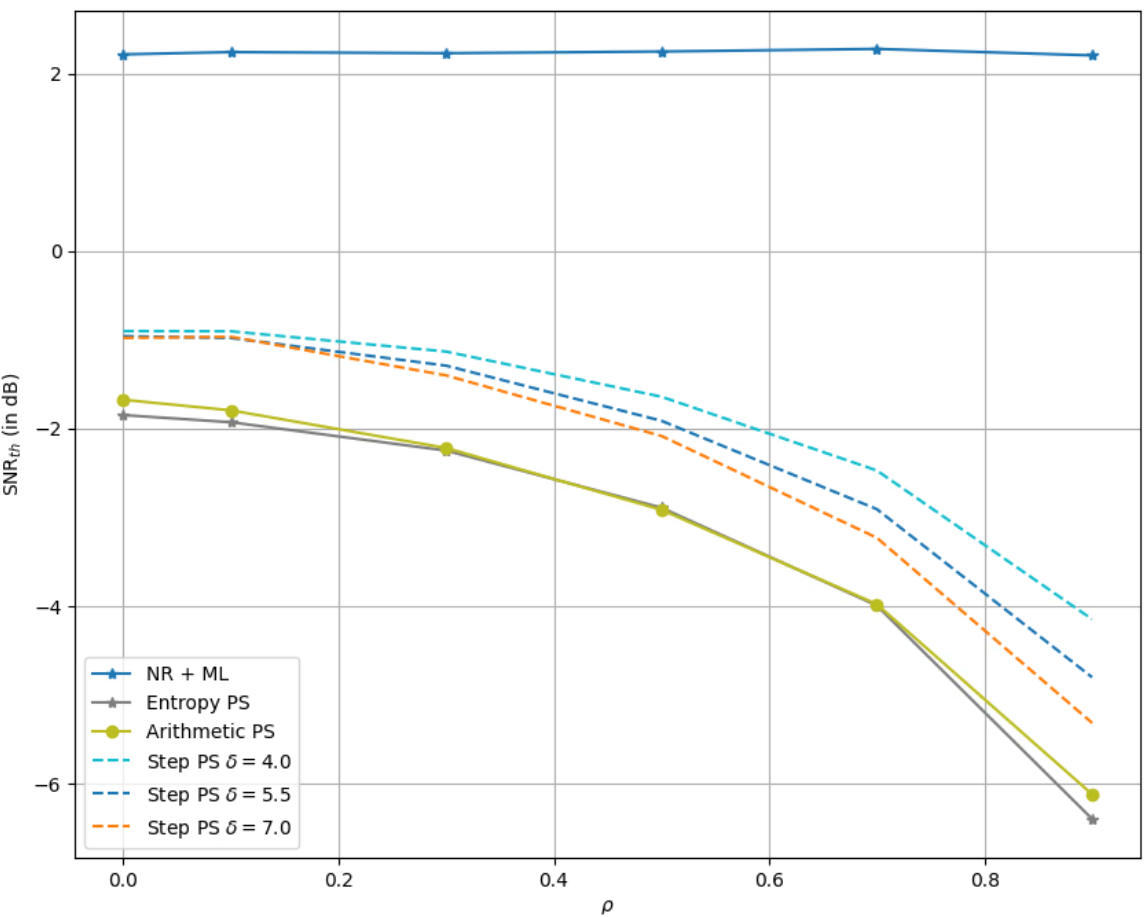}
    \caption{Entropy, Arithmetic and Step PS with FL TF and Bitwise UEP.}	
    \label{fig:ps}
\end{figure}

\begin{figure}
\centering
    \subfloat[$\rho = 0$]{\includegraphics[width = 2in]{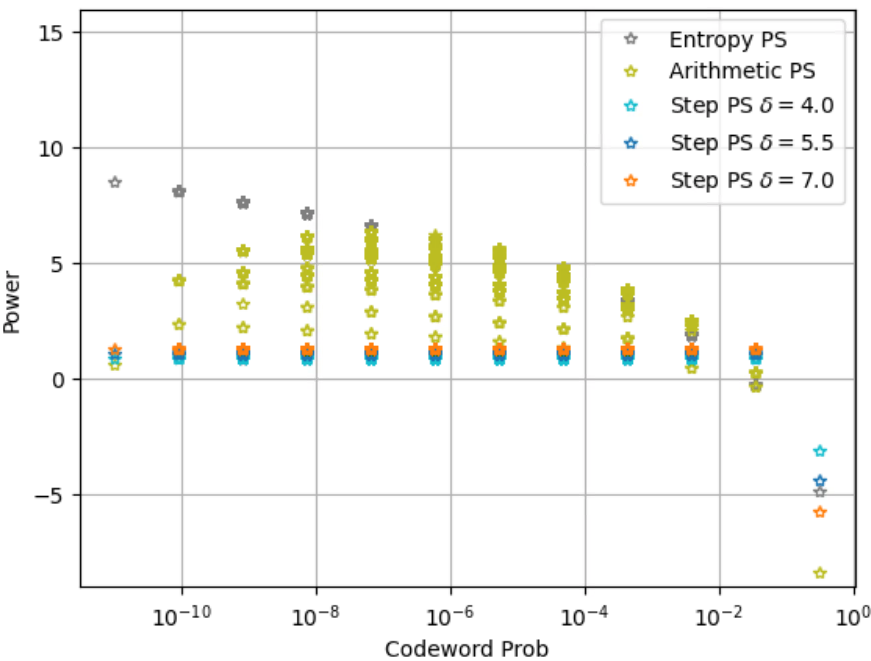}
    \label{fig:rho_0}}
    \hspace{0.01in}
    \subfloat[$\rho = 0.5$]{\includegraphics[width = 2in]{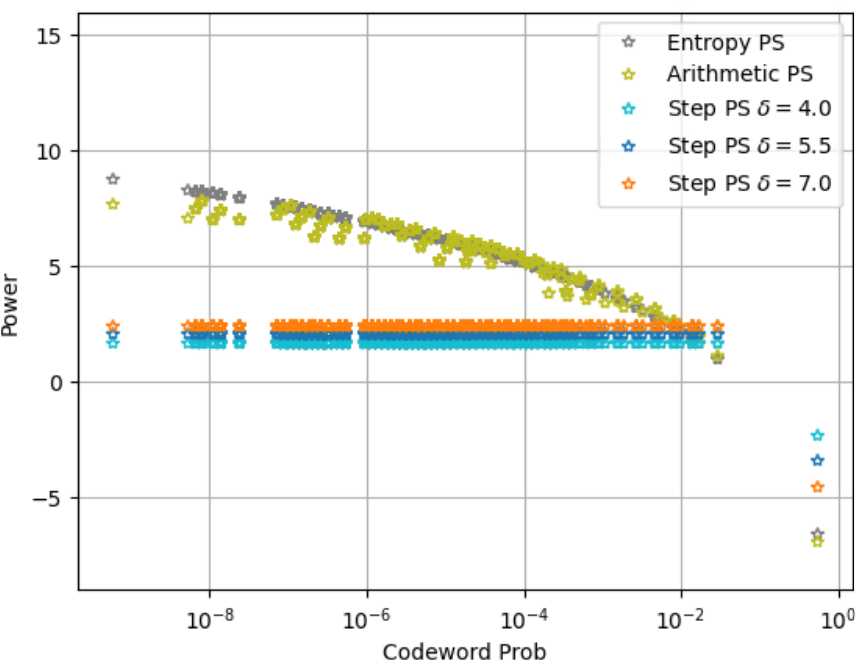}
    \label{fig:rho_0p5}}
    \hspace{0.01in}
    \subfloat[$\rho = 0.9$]{\includegraphics[width=2in]{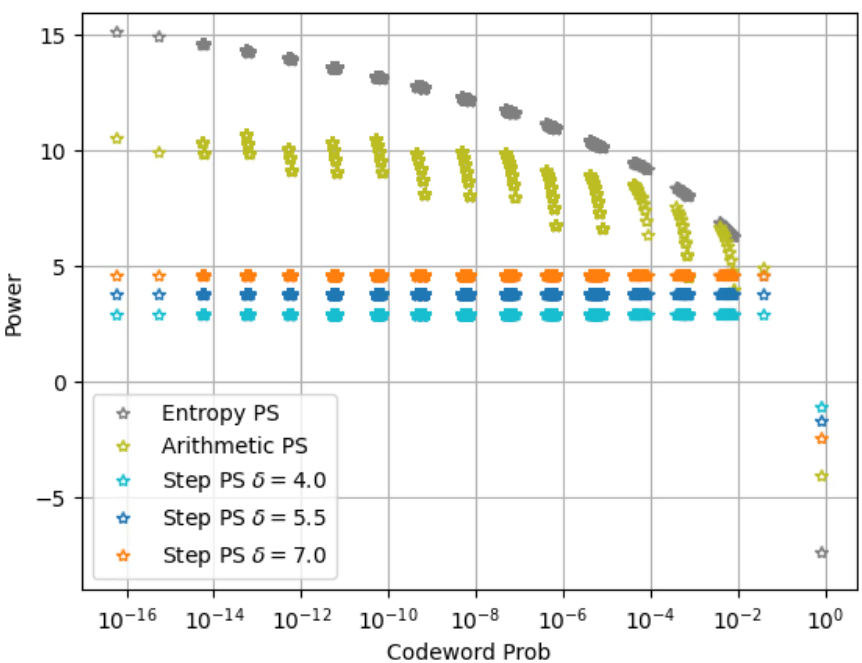}
    \label{fig:rho_0p9}}
\caption{$\alpha_m$ as a function of $\pi_m$.}
\label{fig:plots_ps}
\end{figure}

In order to make the PS scheme less dependent on the exact value of $\mathbf{\Pi}$, we proposed Arithmetic and Step PS. The performance of Entropy, Arithmetic and Step PS is plotted in Fig. \ref{fig:ps}, and the per-codeword power levels for $p = 0.9$ and three different correlation values $\rho = \{0,0.5,0.9\}$ for Entropy, Arithmetic and Step PS are plotted in Fig. \ref{fig:plots_ps}. For Step PS, we have considered three values of $\delta = \{4,5.5,7\}$ dB. While the best performance is provided by Entropy PS, providing gains of $4$ dB for $\rho =0$ and up to 8.6 dB for $\rho=0.9$, it leads to the largest power gap $\Delta = P_{\mathrm{max}} - P_{\mathrm{min}}$ as shown in Fig. \ref{fig:plots_ps} of 13 dB at $\rho = 0$ and 23 dB at $\rho = 0.9$. Arithmetic PS reduces the power gap to 14 dB for all $\rho$, while leading to negligible degradation in performance, with $3.8 - 8.3$ dB of gain over NR + ML. Finally, with Step PS, $\delta = \Delta$ and as we reduce $\delta$, we trade-off performance for reduced power variation. With $\delta = 7$ dB, we achieve $3.2 - 7.5$ dB reduction in $\mathrm{SNR}_{\mathrm{th}}$ while $\delta = 4$ dB yields $3.1 - 6.3$ dB gain. Hence, we can still achieve significant gains over NR + ML with Step PS, while providing for $\sim 7-19$ dB reduction in the maximum power gap compared to Entropy PS.

\subsection{Quantizing TF output}
In Section \ref{subsec:quantization}, we described quantizing the output of the transformer to $\{0,1\}^n$. Having quantized the codebook, the data PAPR of the proposed encoders and the NR baseline would be (nearly) identical per uplink PUCCH transmission. Since uplink transmissions are power limited, it is important to ensure that $P_{\mathrm{max}}$ of the proposed encoders does not exceed that of NR. Assuming Step PS, suppose the all-ACK codeword is transmitted at power $P_0$ and all the other codewords are transmitted at $P_0 + \delta$ dB. Hence, $P_{\mathrm{max}} = P_0 + \delta$ dB. $\mathrm{SNR}_{\mathrm{th}}$ can be interpreted as the average receive power $P_{\mathrm{ave}}$ required to achieve UEP assuming noise power $N_0 = 1$. To ensure that $P_{\mathrm{max}}$ of the proposed encoders does not exceed that of NR, we require $P_{\mathrm{max}} \leq \mathrm{SNR}_{\mathrm{th},NR}$, since NR does not perform PS. 

\begin{figure}
\centering
    \subfloat[$P_{\mathrm{ave}}$ vs $\rho$]{\includegraphics[width = 3.1in]{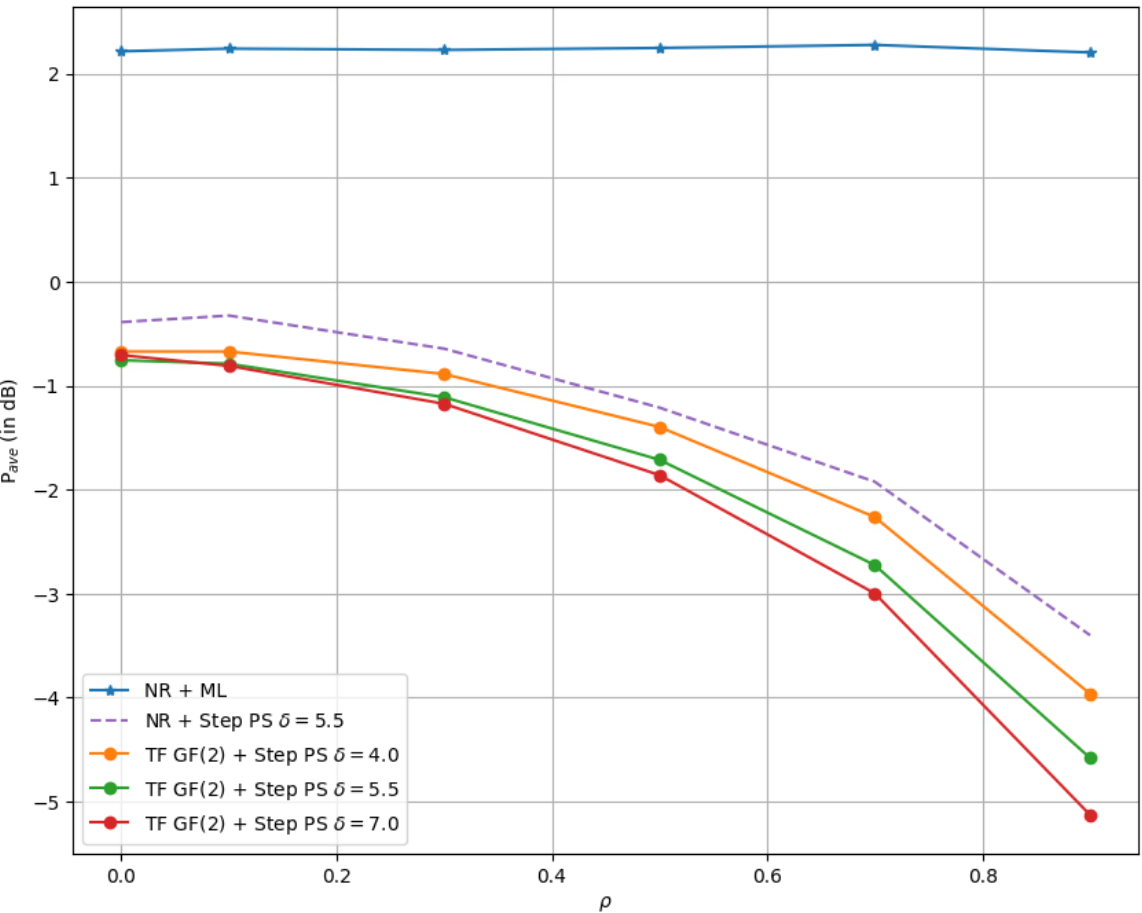}
    \label{fig:p_ave}}
    \hspace{0.01in}
    \subfloat[$P_{\mathrm{max}}$ vs $\rho$]{\includegraphics[width=3.1in]{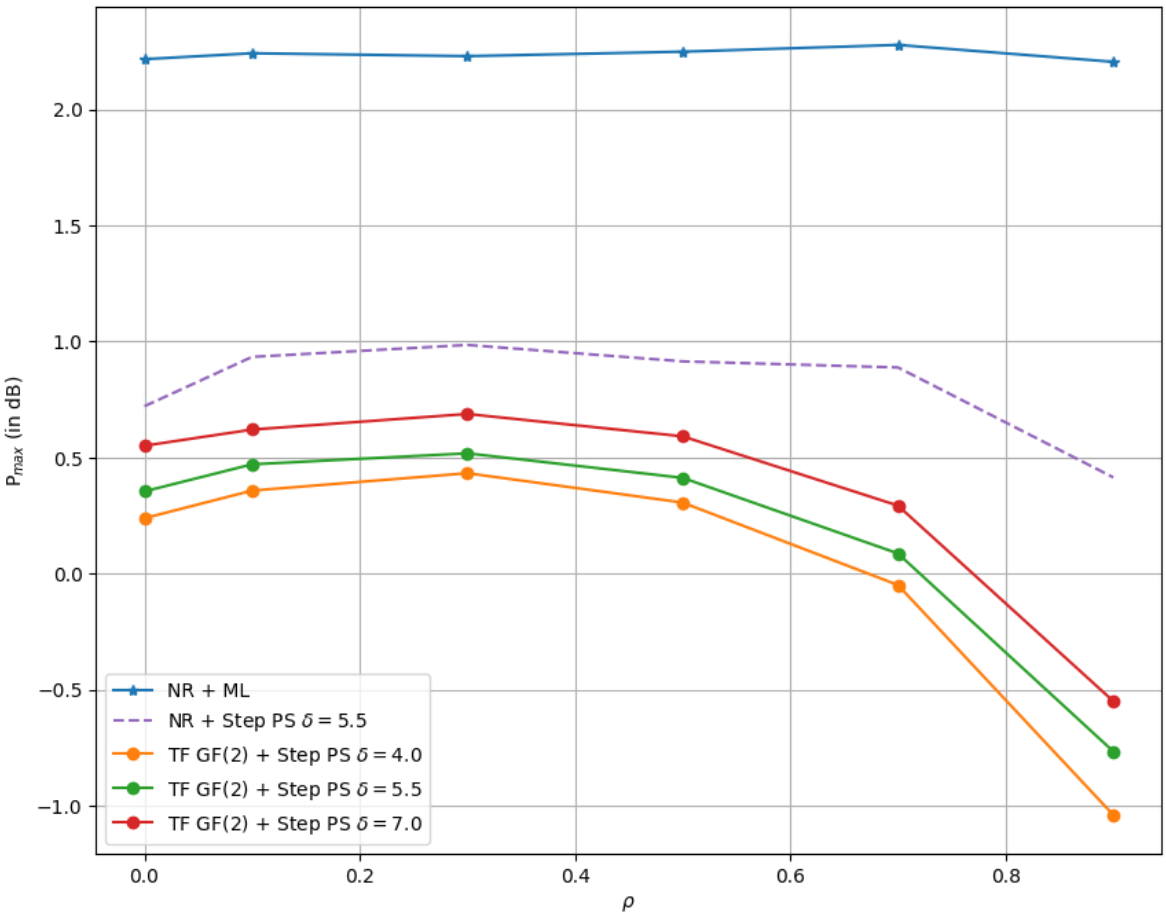}
    \label{fig:p_max}}
\caption{$P_{\mathrm{ave}}$ and $P_{\mathrm{max}}$ vs $\rho$ for GF(2) (binary) TF codebook}
\label{fig:gf_2}
\end{figure}

Fig. \ref{fig:p_ave} and \ref{fig:p_max} plot $P_{\mathrm{ave}} = \mathrm{SNR}_{\mathrm{th}}$ and $P_{\mathrm{max}}$ as a function of $\rho$ respectively for NR + ML decoder, NR + Step PS $\delta = 5.5$ dB + Bitwise UEP and TF + Step PS + Bitwise UEP with $\delta = \{4,5.5,7\}$ dB. We observe that TF + Step PS provides an average reduction in $\mathrm{SNR}_{\mathrm{th}}$ over NR + ML of 3 dB at $\rho = 0$ and up to 6.7 dB at $\rho = 0.9$. More notably, however, both NR and TF when combined with Step PS and UEP provide reductions in $P_{\mathrm{max}}$ as illustrated in Fig. \ref{fig:p_max}, with NR and TF providing 1.3 and 1.8 dB reduction on average respectively. Thus the proposed schemes can provide both coverage gain and average power savings simultaneosuly.

Furthermore, comparing with NR + Step PS $\delta = 5.5$ dB, we see that the quantized TF based codebook alone provides 0.3 dB gain at $\rho = 0$ and up to 1.3 dB at $\rho = 0.9$ . To understand why the gain from the TF codebook alone over NR improves at higher $\rho$, it is instructive to look at the cosine similarity of the all ACK and all NACK codeword, which is defined as
\begin{equation}
    \mathrm{CS}(\mathbf{c}_i, \mathbf{c}_j) = \frac{\mathbf{c}_i^T \mathbf{c}_j}{\|\mathbf{c}_i\|_2 \|\mathbf{c}_j\|_2}.
\end{equation}
For instance, if $\rho = 1$, there will be only two codewords and it is optimum to have them antipodal i.e. the all ACK codeword could be $\mathbf{1}_n$ and the all NACK codeword $-\mathbf{1}_n$, hence the CS is -1. Since FL comprises training over several $\rho$, it cannot learn an antipodal assignment. Nevertheless, the CS between all ACK and all NACK codeword for TF is $-0.81$ compared to NR for which it is $-0.375$. Since TF is closer to antipodal while NR is closer to orthogonal, the larger gains over NR at high $\rho$ are to be expected.

\subsection{Comparing Bitwise UEP and MLM GLRT} \label{subsec:mlm_glrt_results}
\begin{figure}
\centering
    \subfloat[$P_{\mathrm{ave}}$ vs $\rho$]{\includegraphics[width = 3.1in]{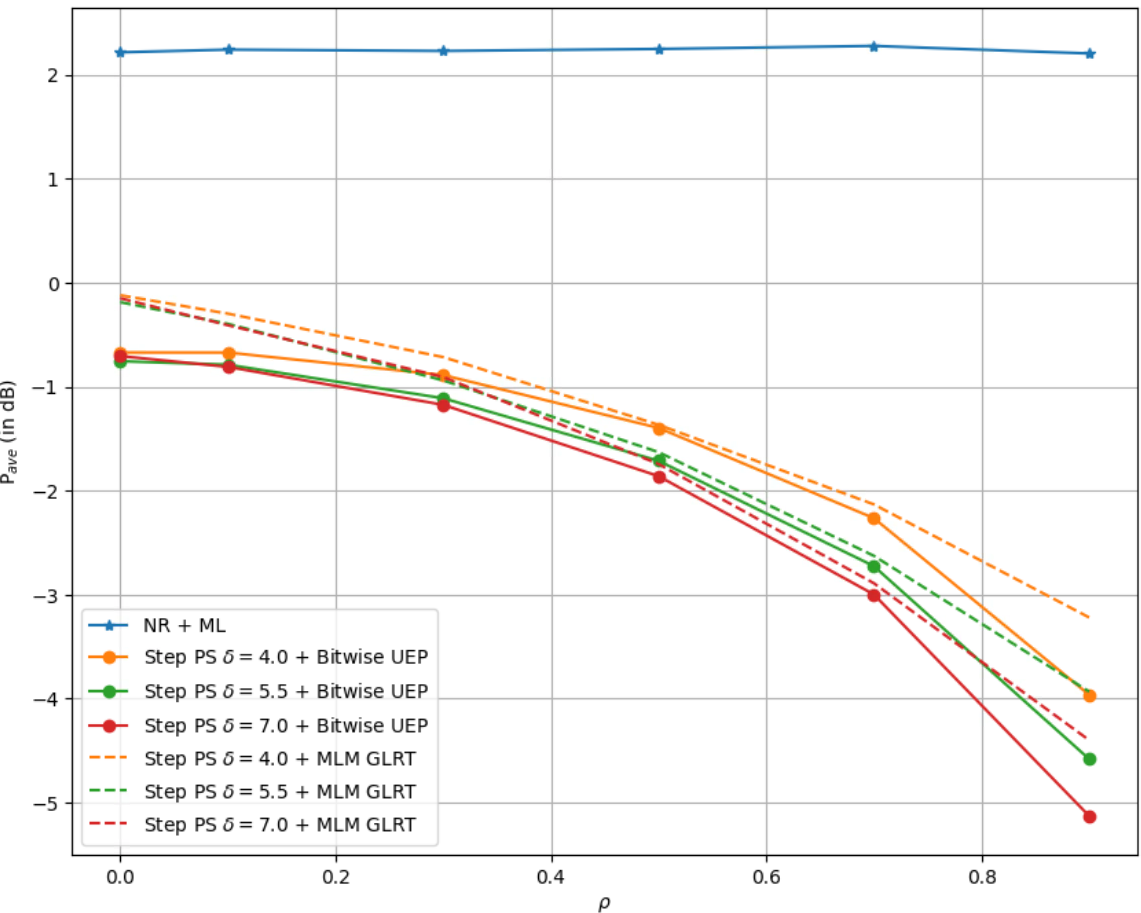}
    \label{fig:p_ave_mlm_glrt}}
    \hspace{0.01in}
    \subfloat[$P_{\mathrm{max}}$ vs $\rho$]{\includegraphics[width=3.1in]{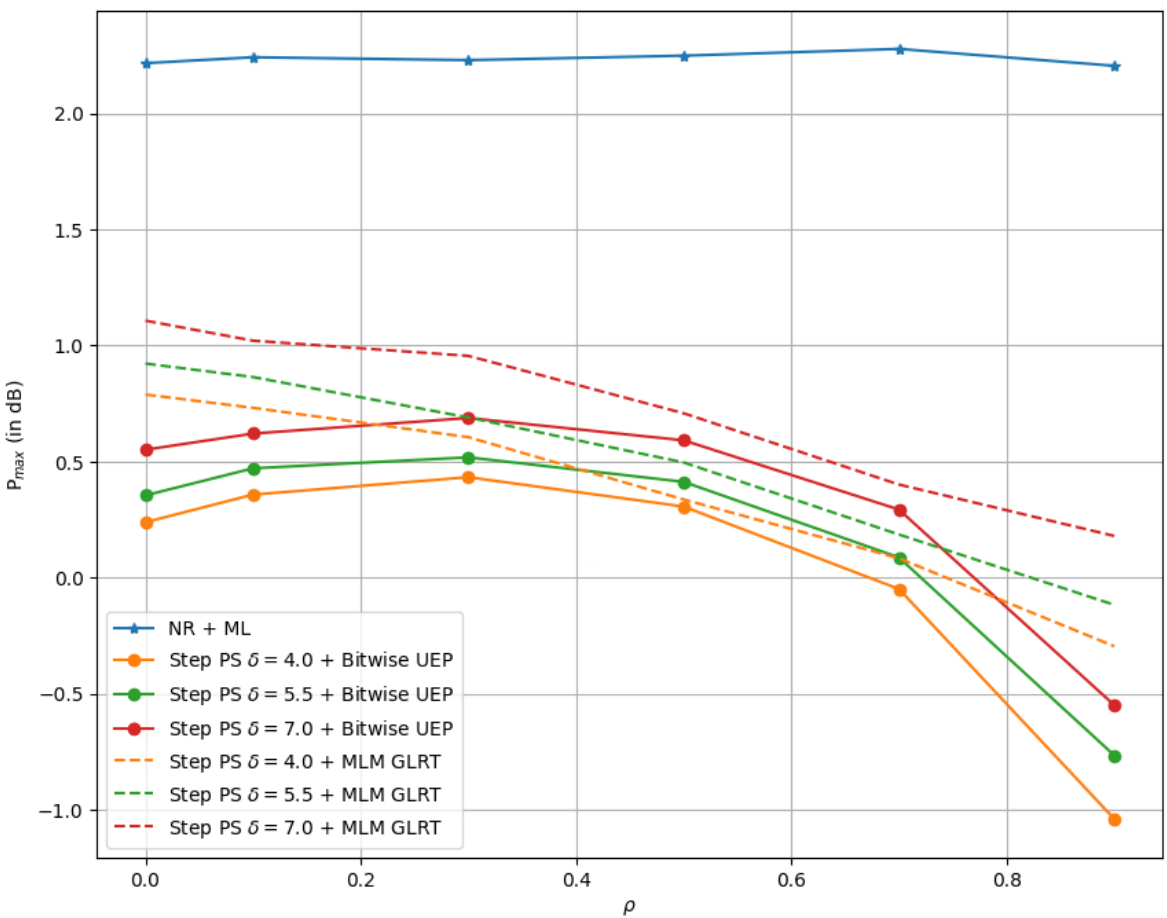}
    \label{fig:p_max_mlm_glrt}}
\caption{Comparing MLM GLRT and Bitwise UEP in term of $P_{\mathrm{ave}}$ and $P_{\mathrm{max}}$ vs $\rho$ for GF(2) (binary) TF codebook}
\label{fig:gf_2_MLM_GLRT}
\end{figure}

Having addressed the practical concerns on the encoder side -- robustness to distributional shifts and lowering PAPR, we proceed to address practical concerns on the decoder side. As introduced in Section \ref{subsec:mlm_glrt}, the BS may have partial knowledge of the source HARQ-ACK distribution. We consider that the BS may be privy to $p$ since it is aware of the PDSCH BLER targeted by OLLA, but it may not know $\rho$, the inter-bit correlation. Hence, \eqref{eq:glrt} reduces to 
\begin{equation} \label{eq:mlm_glrt_ar1}
\frac{\max_{\rho} p_{\rho}(\mathbf{y}|b_i=0)}{\max_{\rho} p_{\rho}p(\mathbf{y}|b_i=1)} \geq \frac{\beta}{1-\beta}.
\end{equation}
Note that the $\rho$ that maximizes $p_{\rho}(\mathbf{y}|b_i=0)$ and $p_{\rho}(\mathbf{y}|b_i=1)$ may be different. Similarly, we can rewrite \eqref{eq:mlm_glrt} by replacing $\Psi_1$ by $\rho$. We draw $\rho \in \{0,0.1,0.3,0.5,0.7,0.9\}$. $P_{\mathrm{ave}} = \mathrm{SNR}_{\mathrm{th}}$ and $P_{\mathrm{max}}$ as a function of $\rho$ for $p = 0.9$ with three different TF + PS combinations are plotted in Fig. \ref{fig:p_ave_mlm_glrt} and \ref{fig:p_max_mlm_glrt} respectively for both Bitwise UEP and MLM GLRT receivers. On average, MLM GLRT leads to $0.33$ dB degradation compared to Bitwise UEP across all $\rho$ and $\delta$. Even with $\delta = 4$ dB, the binary TF codebook obtains a $1-2$ dB reduction in $P_{\mathrm{max}}$ and a $2-5$ dB improvement in $P_{\mathrm{ave}}$. Furthermore, the performance degradation of MLM GLRT could be reduced by utilizing a different $\beta$ in \eqref{eq:mlm_glrt_ar1} for each $\rho$.

\section{Fading Channel Receiver Results} \label{sec:practical_receiver_results}
Taking into account the practical considerations on the encoder side, in accordance with the results presented in Section \ref{sec:awgn_receiver_results}, we will utilize a quantized transformer codebook for $(k,n) = (4,32)$ with Step PS of 3 dB and a Bitwise UEP receiver\footnote{$k=4$ was determined to be the most common value for HARQ-ACK in external discussions with other companies.} . The transformer utilizes $A = 2$ tokens of input length 2 each. We use a relatively low value of Step PS to ensure a low power jump across OFDM symbols, if the DMRS is not power shaped and the DMRS and data are time-division multiplexed on consecutive OFDM symbols. A Bitwise UEP receiver is utilized in place of the more practical MLM GLRT receiver, with the understanding that MLM GLRT will lead to $0.3 - 0.4$ dB degradation for all receivers (excluding the ML baseline). This is because imprecise computations of the codeword likelihoods $p(\mathbf{y}|\mathcal{H}_m)$ will require the utilization of different $\beta$ values for different algorithms, which could be hard to analyze with MLM GLRT also being utilized.

\subsection{NR Compliant Uplink Transmission}
\begin{figure}
    \centering
    \includegraphics[width =3.7in]{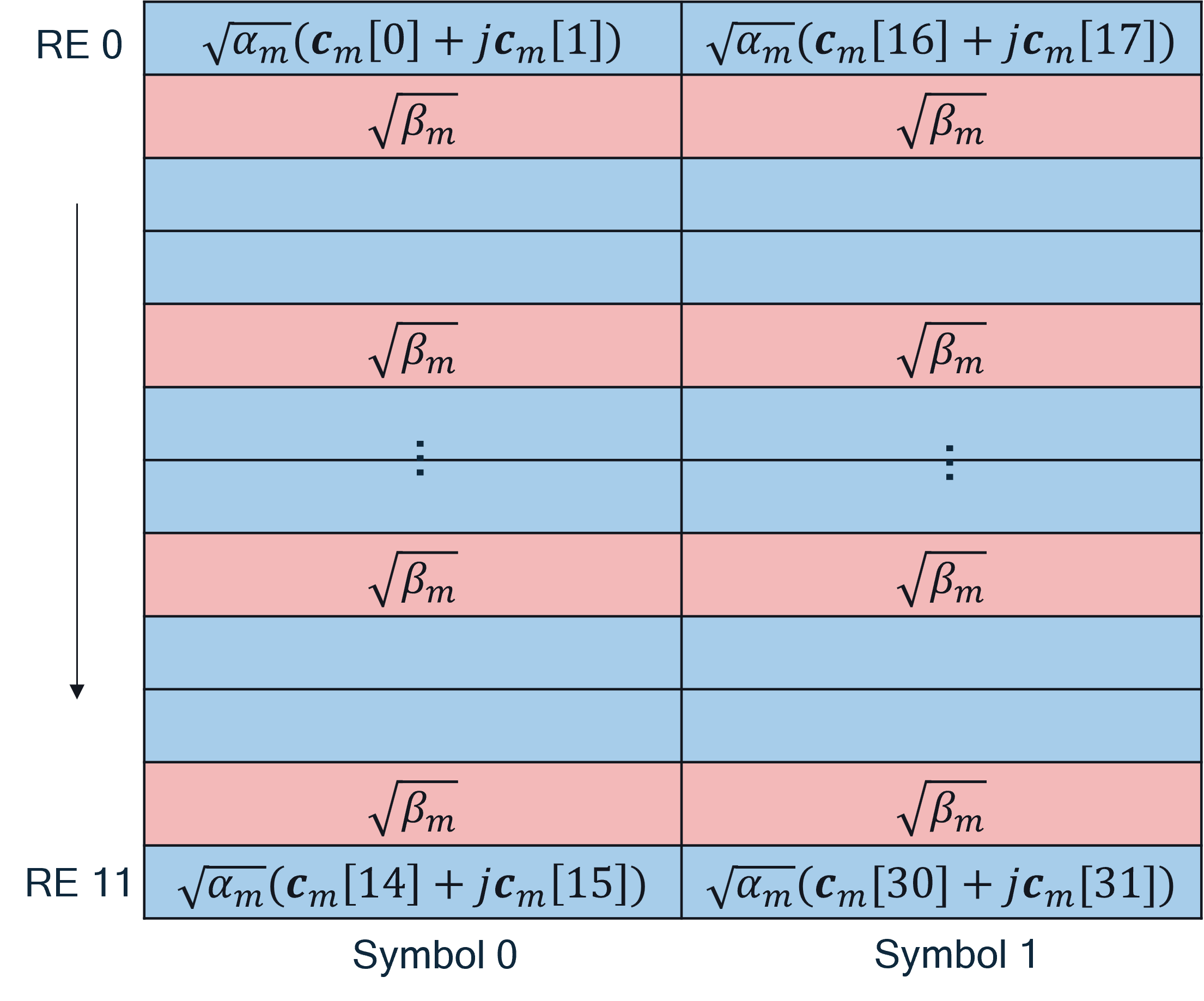}
    \caption{PUCCH Format 2 showing placement and value of data and DMRS REs}
    \label{fig:format_2}
\end{figure}

Consider a SIMO setup with $N_r = 4$ receive antennas. We simulate PUCCH format 2 \cite{3gpp.38.211} over 1 RB and 2 OFDM symbols. The frame structure is depicted in Fig. \ref{fig:format_2}. The transmission consists of 24 REs, with 8 DMRS REs (shown in red) and 16 data REs (shown in blue). Since PUCCH Format 2 is modulated using QPSK, $n = 32$ coded bits are mapped to 16 QPSK symbols and the need for rate matching is alleviated.  The QPSK symbols are scaled by $\sqrt{\alpha_m}$ and mapped to the 16 data REs on the OFDM grid. The DMRS symbols are set to 1, scaled by $\sqrt{\beta_m}$ and mapped to the 8 DMRS REs on the OFDM grid.

\vspace{2mm}
\begin{remark}
\textit{In NR, QPSK is transmitted using OFDM waveform while $\pi/2$-BPSK is transmitted using DFT-s-OFDM in uplink. While we assume QPSK modulation and hence directly place the coded symbols onto the OFDM grid (without transform precoding \cite{3gpp.38.211}), it is possible that 6G and beyond could use DFT-s-OFDM with other modulation schemes, hence we restrict our evaluations to the quantized transformer codebook to ensure low PAPR.}
\end{remark}

\subsection{Computing receive SNR and noise variance in practice} \label{subsec:snr_ne_comp}
While we can assume genie knowledge of $\mathrm{SNR}$ and $\mathrm{NV}$ in simulation, these will have to be computed from reference signals in practice. For computing the SNR, we have two options: it can be estimated from the DMRS REs, or we can assume a nominal value. Given only 8 DMRS REs, the SNR estimated from the DMRS may have high variance. Hence we assume a fixed nominal SNR at the receiver (irrespective of the actual received SNR). In Appendix \ref{subsec:nominal_snr}, we will present empirical evidence that demonstrates minimal degradation from overestimation of the received SNR. Since PUCCH performance gains are to be obtained at low SNR i.e. $< 0$ dB, we will fix the nominal SNR to $\mathrm{SNR}_{\mathrm{n}} = 0$ dB. Given that the DMRS is shaped, for hypothesis $\mathcal{H}_m$, the hypothesized nominal SNR is $\beta_m \mathrm{SNR}_{\mathrm{n}}$. Furthermore, per receive antenna, we can estimate $\mathbb{E}[\mathbf{n}_l^2]$ as $\|\mathbf{y}_{p,l} - \mathbf{\hat{h}}_{p,l} \odot \mathbf{c}_{p,l}\|^2_2/ N_p$, and then average $\mathbb{E}[\mathbf{n}_l^2]$ over all receive antennas to obtain $\mathrm{NV}$. 

\subsection{Channel Estimation}
We will be required to perform ChEST as part of receiver processing in case of the coherent receiver designs. As derived in Section \ref{subsec:optimal_coherent}, we can express $\boldsymbol{\mu}_{d,l}$ in \eqref{eq:mu_sigma} as an elementwise product of the LMMSE channel estimate $\mathbf{W}_m \mathbf{y}_{p,l}$ and the hypothesized codeword $\mathbf{c}_m$. Since the LMMSE channel estimate assumes knowledge of $\mathbf{R}_{\mathbf{hh}}$, which is obtained from channel statistics, it is not realizable in practice. Hence, we employ Robust MMSE (RMMSE) channel estimation \cite{li2002robust} which assumes a uniform power delay profile (PDP) between $0$ and $\tau_{ds,est}$ to estimate $\mathbf{R}_{\mathbf{hh}}$, where $\tau_{ds,est}$ can be estimated from reference signals. For our simulations, we employ the TDL-C \cite{3gpp.38.901} channel model with a RMS delay spread of 300 ns. In RMMSE ChEST, we set $\tau_{ds,est} = 500$ ns to filter out the noise taps since the operating SNR is low. 

Since the DMRS is shaped, the $\mathrm{SNR}$ used for ChEST is $\beta_m \mathrm{SNR}_{\mathrm{n}}$ for hypothesis $\mathcal{H}_m$. As discussed in Section \ref{subsec:snr_ne_comp}, minimal degradation is incurred from overestimating the SNR. Given that we utilize Step PS, we use $\beta_{-1} \mathrm{SNR}_{\mathrm{n}}$ as the ChEST SNR, where $\beta_{-1}$ is the DMRS power scaling associated with the all-NACK codeword, in order to consistently overestimate the SNR.

\vspace{2mm}
\begin{remark}
\textit{Since we are evaluating over 1 RB, we can also consider the channel to be approximately frequency flat and set all entries of $\mathbf{R}_{\mathbf{hh}}$ to be 1 while performing ChEST.}
\end{remark}

\subsection{Comparing coherent and noncoherent receiver designs} \label{subsec:results_prac_coh_noncoh}
In Fig. \ref{fig:practical_receiver}, we plot $\mathrm{SNR}_\mathrm{th}$ vs $\rho$ for various schemes under a fading channel (TDL-C 300 ns). The NR + ML baseline is a correlator based decoder i.e. we compute $\Re(\sum_{l} \mathbf{y}^H_{d,l} (\mathbf{\hat{h}}_{d,l} \odot \mathbf{c}_m))$ for each hypothesized codeword $\mathbf{c}_m$ and find the codeword that maximizes the correlation. Since the ML decoder is unimpacted by $\rho$, $\mathrm{SNR}_\mathrm{th}$ is around 0.2 dB for all $\rho$.  This is the coherent ML decoder that would be utilized in practice. However, the optimum codeword ML decoder would be noncoherent and involve computing $p(\mathbf{y}_l | \mathcal{H}_m)$ using \eqref{eq:codeword_likelihood} and then finding the codeword index $m^* = \arg \max_{m} p(\mathbf{Y}|\mathcal{H}_m)$. This improves $\mathrm{SNR}_\mathrm{th}$ by 2 dB. If we utilize the Bitwise UEP receiver in place of the codeword ML decoder, having computed  $p(\mathbf{y}_l | \mathcal{H}_m)$ optimally, a further 1 dB improvement in $\mathrm{SNR}_\mathrm{th}$ is obtained. Finally, a 1 dB improvement can be obtained from Step PS with $\delta = 3$ dB and $0.5-0.7$ dB gain can be obtained from employing the TF-based encoder. Note that $\delta = 3$ dB is a very pessimistic value, and further gains can be obtained from PS by increasing $\delta$, while ensuring $P_{\mathrm{max}}$ is less than $\mathrm{SNR}_\mathrm{th}$ of the NR + ML baseline. Since the optimal coherent and noncoherent decoder are theoretically identical\footnote{As derived in Section \ref{subsec:optimal_coherent}, if we utilize an LMMSE channel estimator whilst utilizing the optimal coherent decoder, then $p(\mathbf{y}_l|\mathcal{H}_m)$ computed in Section \ref{subsec:optimal_coherent} and \ref{subsec:optimal_noncoherent} are identical.}, we confirm via simulation that the $\mathrm{SNR}_\mathrm{th}$ obtained from both line up. Replacing the optimal coherent decoder by its low complexity approximation leads to a minimal degradation of $0.3$ dB for TF and NR. 

\subsection{Impact of DMRS Power Shaping} \label{subsec:dmrs_ps_results}
\begin{figure}
\centering
    \subfloat[Comparing Optimal Noncoherent, Coherent and Low Complexity Coherent Receiver Design.]{\includegraphics[width = 3.1in]{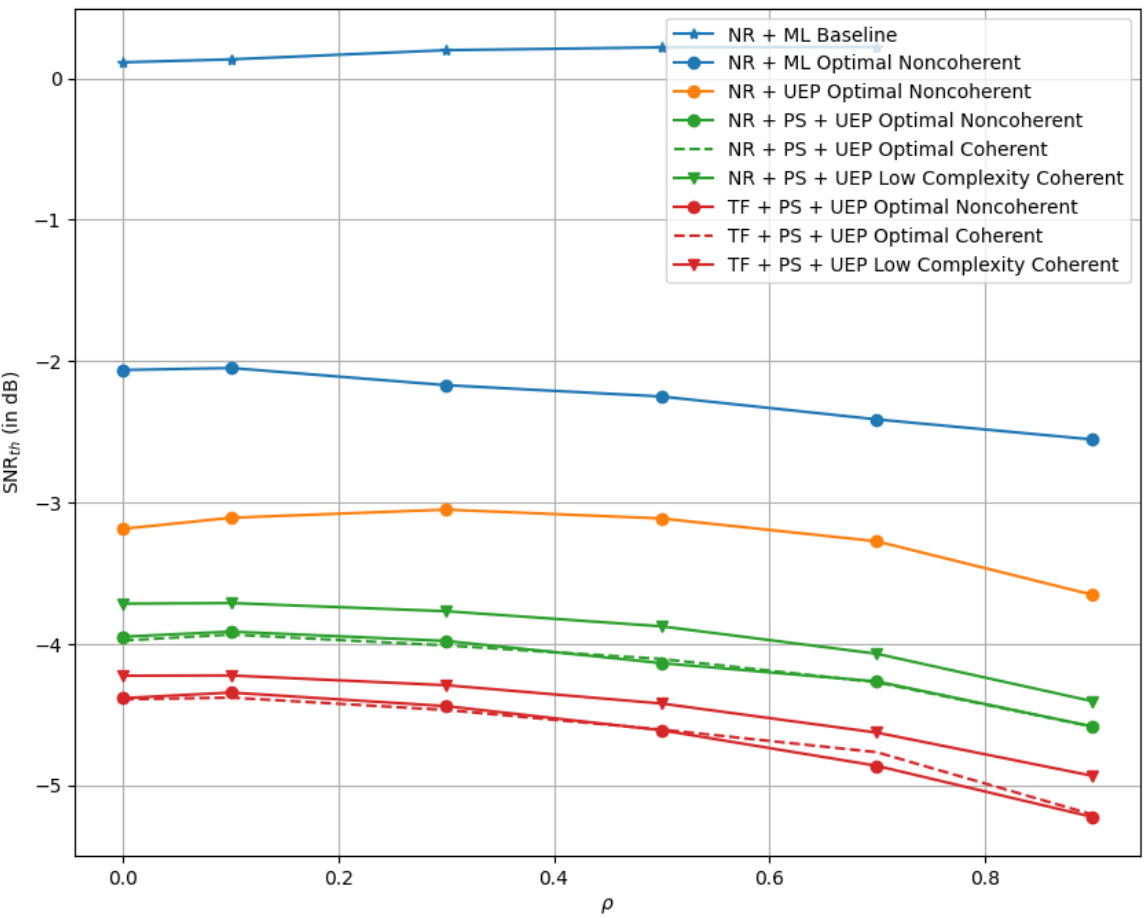}
    \label{fig:practical_receiver}}
    \hspace{0.01in}
    \subfloat[Impact of DMRS power shaping on Optimal Noncoherent Receiver performance.]{\includegraphics[width=3.1in]{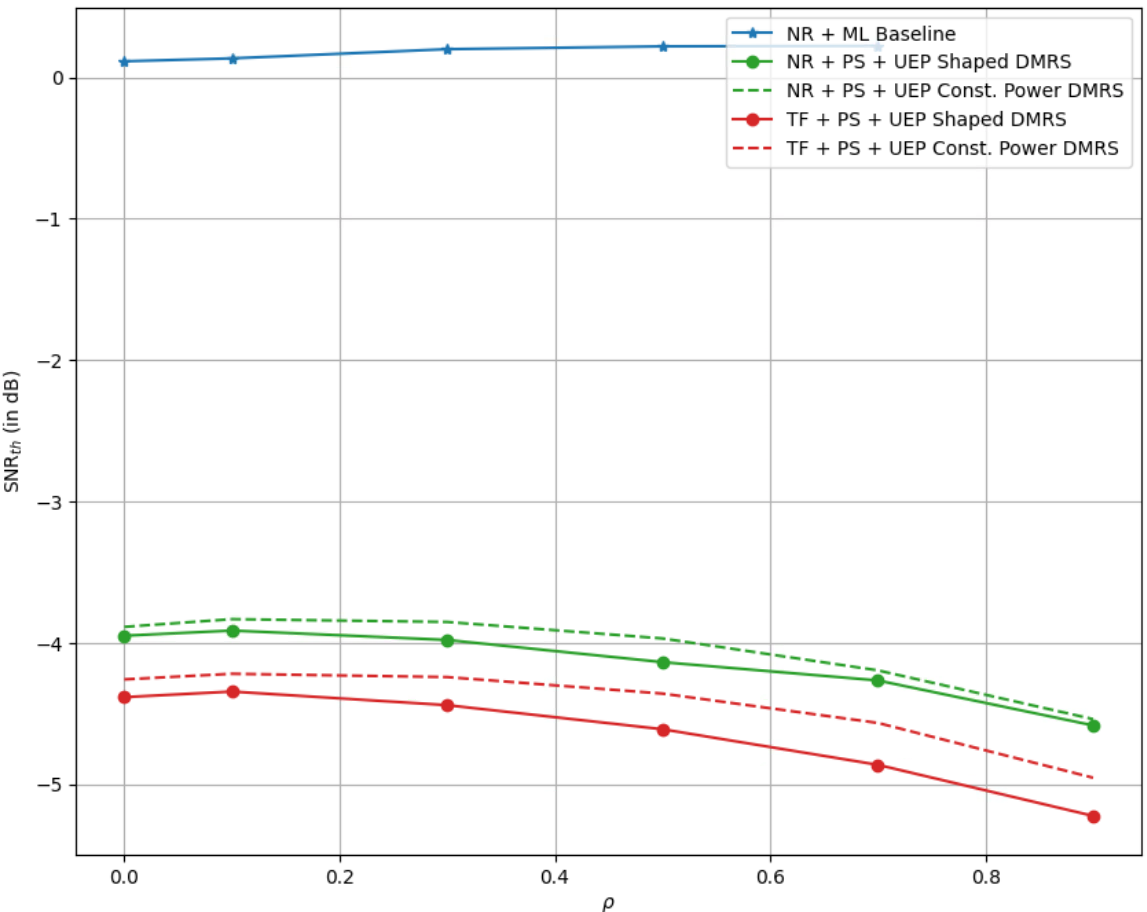}
    \label{fig:practical_receiver_dmrs}}
\caption{Fading Receiver Design Results}
\label{fig:prac_results}
\end{figure}

In Fig. \ref{fig:practical_receiver_dmrs}, we investigate the impact of setting $\beta_m = 1$ (Const. Power DMRS) vs $\beta_m = \alpha_m$ (Shaped DMRS). We observe that utilizing a constant power DMRS over the shaped DMRS leads to a $\sim 0.3$ dB degradation for TF + Step PS and $\sim 0.1$ dB degradation for NR + Step PS. By shaping the DMRS, we are effectively increasing the separation between the (data + DMRS) codewords $\mathbf{c}_{p+d,m}$, and hence the shaped DMRS scheme outperforms. Furthermore, shaping the DMRS power reduces the power variation in the transmitted signal and prevents RF issues.

\section{Conclusions and Future Directions} \label{sec:conclusions}
Since NR channel codes have been designed assuming a uniform source prior, they do not jointly perform source encoding that could exploit the biased HARQ-ACK source prior which arises from the PDSCH BLER being maintained under $10\%$\footnote{Even with the current NR codebook, the decoder could always utilize the HARQ-ACK prior to provide power savings.}. In this paper, we make fundamental contributions to joint source and channel coding (and modulation) for HARQ-ACK bits transmitted in the uplink. On the encoder side, we propose a non-linear transformer-based codebook to exploit the source prior. Furthermore, we demonstrate how assigning less power to more frequently seen codewords and vice-versa can lower the SNR required to achieve the target error rates. On the decoder side, we propose the first extension of the Neyman Pearson test to a coded bit system with multiple information bits to preferentially protect NACK over ACK. This further lowers the SNR needed to achieve a NACK and ACK error rate of 0.1\% and 1\% respectively. For both the encoder and decoder, we propose methods to lower complexity and improve robustness to small changes and/or partial knowledge of the HARQ-ACK distribution. Finally, we present comprehensive end-to-end NR-compliant simulations in an uplink SIMO setup and develop the receiver from first principles to demonstrate the effectiveness of JSCC for HARQ-ACK. By obtaining 3 -- 6 dB reduction in average transmit power and 2 -- 3 dB reduction in the maximum transmit power over the NR baseline while accounting for practical constraints, our approach provides for significant power savings and coverage gains.

As part of future work, a single codebook can be designed that works for a range of $(k,n)$ values, akin to the generator matrix used in \cite{3gpp.38.212}. While we presented results for two $k$ values, $k=4$ and $k=11$, and designed a separate transformer for each, having a single codebook for $3 \leq k \leq 11$ would be desirable. For larger payloads $k > 11$, an ML/MAP decoder would become computationally infeasible, hence an AI based encoder would have to be coupled with an AI based decoder to ensure manageable computational complexity. On a theoretical front, proving the optimality of $\beta$ as defined in \eqref{eq:thresh} remains an open question. We hope that the first principles based approach taken in this paper can inspire further innovation in the development of practical AI/ML based enhancements to the PHY layer for 6G and beyond.

\section{Acknowledgments}
The authors would like to thank Yi Huang, Peter Gaal, Mike McCloud, Brian Banister and Hari Sankar for their valuable inputs and constructive feedback over the course of this project. We would also like to sincerely thank the reviewer who took the effort to provide in entirety the correct derivation for Appendix \ref{subsec:one_bit_uep}.

\bibliographystyle{IEEEtran}
\bibliography{bibtex.bib}

\appendix
\renewcommand{\thesubsection}{\Alph{subsection}}

{\subsection{Group Permutation Symmetry} \label{subsec:perm_equi}
Let $\mathbf{X} = \{\mathbf{x}_1, \mathbf{x}_2, \ldots, \mathbf{x}_n\}$ be a collection of input sequences, and let $\pi$ be any permutation of the indices $\{1, \ldots, n\}$. A function $f: \mathcal{X}^n \rightarrow \mathcal{Y}^n$ is said to be permutation equivariant if, for every permutation $\pi$, 
\begin{equation}
    f(\pi\mathbf{X}) = \pi f(\mathbf{X}).
\end{equation}
For example, consider the transformer architecture depicted in Fig. \ref{fig:transformer}. It has been shown that transformers without positional embedding are permutation equivariant w.r.t token permutations (refer Theorem 4.1 in \cite{xu2024permutation}). This implies that given a set of $A$ input tokens $\mathbf{x}$ and the transformer output for this sequence $H(\mathbf{x})$, the transformer output for any token permutation of $\mathbf{x}$ can be obtained by permuting the output tokens contained in $H(\mathbf{x})$. We refer to this property as ``group permutation symmetry". If we increase $A$, a greater number of input sequences can be represented as a token permutation of a given input sequence $\mathbf{x}$, thus leading to an increased degree of group permutation symmetry.

In the absence of the said symmetry, the size of the LUT required to store the transformer output would be $(2^k,n)$, $n$ real values for each of the $2^k$ possible input sequences. However, if we do have group permutation symmetry with $A$ tokens of size $\lceil k/A \rceil$, then the size of the LUT is
\begin{equation}
    \bigg(2^{\lceil k/A \rceil} \binom{2^{\lceil k/A \rceil} + A - 2}{2^{\lceil k/A \rceil} - 1}, \lceil n/A \rceil\bigg).
\end{equation}
This follows since we have $2^{\lceil k/A \rceil}$ possible input tokens and for each input token, we have an output sequence of length $\lceil n/A \rceil$. Furthermore, the transformer encoder determining the output sequence for the given input token is token permutation \textit{in}-variant to the other $A-1$ input tokens. If we count the number of distinct $A-1$ input token sequences, ignoring any token permutations, where each token can take $2^{\lceil k/A \rceil}$ distinct values, then that can easily be shown to be $\binom{2^{\lceil k/A \rceil} + A - 2}{2^{\lceil k/A \rceil} - 1}$.

Consider $k = 11$, $n = 32$ and $A = 4$. Then the reduced size of the LUT is $(960,8)$ which is smaller than $(2^{11},32)$ by a factor of 8.5.

\subsection{Learnt PS - Training and Results} \label{subsec:learnt_ps_appendix}
In Section \ref{sec:enc_design}, the output of the transformer-based encoder is normalized to ensure that $\|\mathbf{c}_m\|^2_2 = n/2$ $\forall m$. In order to learn both the power shaping and codebook jointly, we could choose not to normalize the output as briefly mentioned in Section \ref{subsec:problem_statement}. In other words, let us denote the output of the transformer prior to normalization as $\mathbf{c}_{u,m}$ (assume no quantization). We input all the message sequences $m \in [2^k]$ to the transformer to obtain the (unnormalized) codebook $\mathcal{C}_u = \{\mathbf{c}_{u,m}\}_{m}$. Given prior $\mathbf{\Pi}$, we then scale each codeword by $\sqrt{\kappa_{\mathbf{\Pi}}}$ such that
\begin{equation}
	\kappa_{\mathbf{\Pi}} \sum_{m \in [2^k]} \pi_m \|\mathbf{c}_{u,m}\|^2_2 = n/2.
\end{equation}
We then replace $\sqrt{\alpha_m} \mathbf{c}_m$ in Algorithm \ref{alg:free_lunch_training} by $\sqrt{\kappa_{\mathbf{\Pi}}} \mathbf{c}_{u,m}$, and the remainder of the training process remains unchanged. If we set $N=1$ and train on $(p,\rho) = (0.9,0)$ (where $\rho \in [0,1]$ is the correlation coefficient between consecutive bits $b_{i-1}$ and $b_i$), we will converge to the Learnt PS depicted in Fig. \ref{fig:entropy_ps}. If we perform free-lunch training as described in Section \ref{subsec:training_algo}, the PS learnt (labeled Robust PS) is very similar to Arithmetic PS and is depicted in Fig. \ref{fig:plots_ps_robust} for $p=0.9$ and $\rho = \{0,0.5,0.9\}$.
\begin{figure}
\centering
    \subfloat[$\rho = 0$]{\includegraphics[width = 2in]{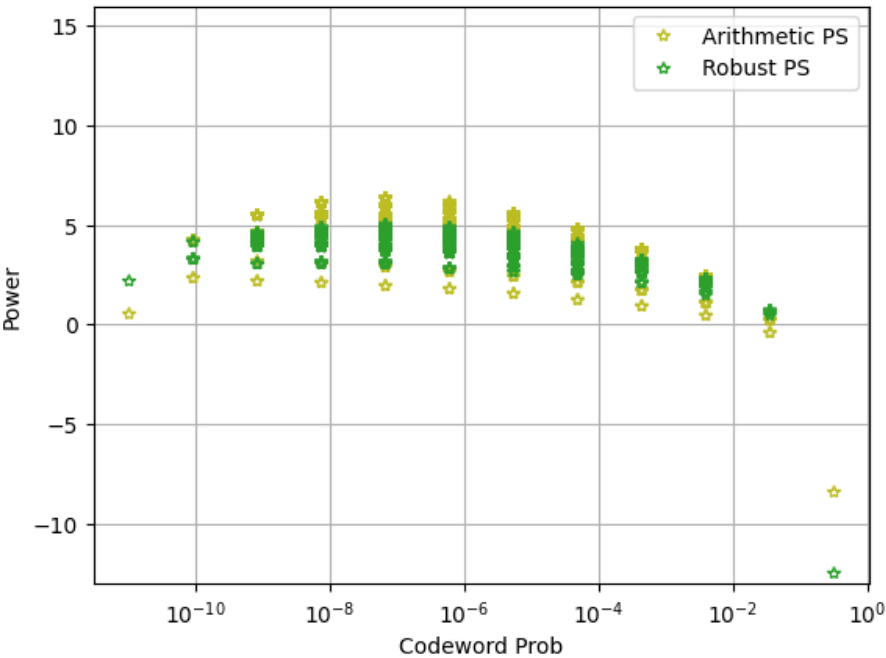}
    \label{fig:rho_0_robust}}
    \hspace{0.01in}
    \subfloat[$\rho = 0.5$]{\includegraphics[width = 2in]{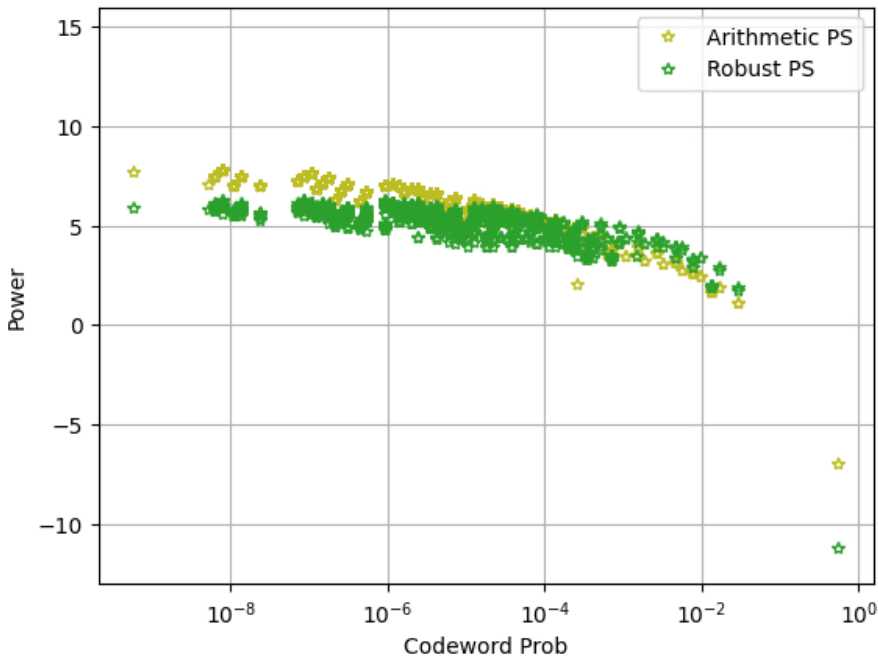}
    \label{fig:rho_0p5_robust}}
    \hspace{0.01in}
    \subfloat[$\rho = 0.9$]{\includegraphics[width=2in]{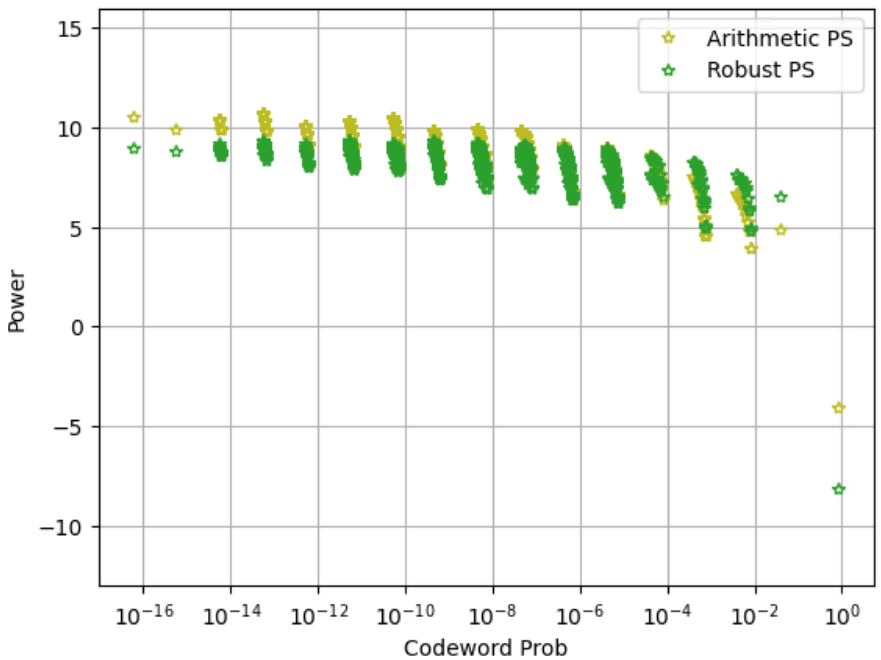}
    \label{fig:rho_0p9_robust}}
\caption{$\alpha_m$ as a function of $\pi_m$ for Robust and Arithmetic PS.}
\label{fig:plots_ps_robust}
\end{figure}

\subsection{Bitwise UEP for 1 bit} \label{subsec:one_bit_uep}
Consider the vector complex AWGN channel
\begin{equation}
 \mathbf{y}=\mathbf{x}_i+\mathbf{n},\qquad \mathbf{n} \sim  \mathcal{CN}(0,\sigma^2\mathbf{I}_{n/2}),
\end{equation}
so that $\mathbb{E}[\lVert \mathbf{n} \rVert^2]=\sigma^2n/2$ and equivalently $\Re\{\mathbf{n}\},\Im\{\mathbf{n}\}\sim\mathcal{N}(0,\frac{\sigma^2}{2} \mathbf{I}_{n/2})$. Assume binary signaling for the HARQ-ACK bit with repetition coding:
\begin{equation}
 \mathbf{x}_1=+\sqrt{P_1} \mathbf{1}_{n/2}\ \text{(ACK)},\qquad \mathbf{x}_0=-\sqrt{P_0} \mathbf{1}_{n/2}\ \text{(NACK)}.
\end{equation}
The conditional pdf of $\mathbf{y}$ given $\mathbf{x}_i$ is
\begin{equation}
 p(\mathbf{y}\mid \mathbf{x}_i)=\frac{1}{\pi^{n/2}\sigma^2}\exp\!\left(-\frac{\lVert \mathbf{y}-\mathbf{x}_i\rVert^2}{\sigma^2}\right).
\end{equation}
A likelihood-ratio test of the form
\begin{equation}
\log\frac{p(\mathbf{y}\mid \mathbf{x}_0)}{p(\mathbf{y}\mid \mathbf{x}_1)}\ge \log\frac{\beta}{1-\beta}
\end{equation}
is equivalent to
\begin{equation} \label{eq:app_1}
\lVert \mathbf{y}-\mathbf{x}_1\rVert^2-\lVert \mathbf{y}-\mathbf{x}_0\rVert^2 \ge \sigma^2\log\frac{\beta}{1-\beta}.
\end{equation}
Expanding the difference of squared distances yields
\begin{equation} \label{eq:app_2}
\begin{aligned} 
\lVert \mathbf{y}-\mathbf{x}_1\rVert^2-\lVert \mathbf{y}-\mathbf{x}_0\rVert^2
&=\big(\lVert \mathbf{y}\rVert^2-2\Re\{\mathbf{x}_1^H\mathbf{y}\}+\lVert \mathbf{x}_1\rVert^2\big)
 -\big(\lVert \mathbf{y}\rVert^2-2\Re\{\mathbf{x}_0^Hy\}+\lVert \mathbf{x}_0\rVert^2\big)\\
&=-2\Re\{(\mathbf{x}_1-\mathbf{x}_0)^*y\}+\big(\lVert \mathbf{x}_1\rVert^2-\lVert \mathbf{x}_0\rVert^2\big).
\end{aligned}  
\end{equation}
Let $\mathbf{d}\triangleq \mathbf{x}_1-\mathbf{x}_0$. Substituting \eqref{eq:app_2} into \eqref{eq:app_1} gives the linear test
\begin{equation}
\Re\{\mathbf{d}^H\mathbf{y}\}\le \tau,\qquad
\tau\triangleq \frac{\lVert \mathbf{x}_1\rVert^2-\lVert \mathbf{x}_0\rVert^2-\sigma^2\log\frac{\beta}{1-\beta}}{2}.
\end{equation}
For $\mathbf{x}_1=+\sqrt{P_1} \mathbf{1}_{n/2}$ and $\mathbf{x}_0=-\sqrt{P_0} \mathbf{1}_{n/2}$, we have $\lVert \mathbf{x}_1\rVert^2-\lVert \mathbf{x}_0\rVert^2=n/2(P_1-P_0)$, and hence $\Re\{\mathbf{d}^H\mathbf{y}\}=(\sqrt{P_1}+\sqrt{P_0})\Re\{\mathbf{1}^H\mathbf{y}\}$. Therefore (3) reduces to $\Re\{\mathbf{1}^H\mathbf{y}\}\le\gamma$, where
\begin{equation}
\gamma\triangleq \frac{n/2(P_1-P_0)-\sigma^2\log\frac{\beta}{1-\beta}}{2(\sqrt{P_1}+\sqrt{P_0})}.
\end{equation}
Equivalently, decide ACK iff $\Re\{\mathbf{1}^H\mathbf{y}\}>\gamma$. Under NACK($\mathbf{x}_0=-\sqrt{P_0} \mathbf{1}_{n/2}$), $\Re\{y\}=-\sqrt{P_0}\mathbf{1}_{n/2}+\Re\{\mathbf{n}\}$, and thus the NACK error rate is given by
\begin{equation} \label{eq:1_bit_v2}
P(\mathrm{ACK}\mid \mathrm{NACK})=P\bigg(\Re\{\mathbf{1}^H\mathbf{y}\}>\gamma \mid \mathbf{x}_0\bigg)=Q\!\left(\frac{\gamma+(n/2)\sqrt{P_0}}{(n/2)\sigma}\right).   
\end{equation}
Given a target NACK error rate $\delta_0$, $\beta$ can be computed as a function of $\sigma^2$ and $\delta_0$.

\subsection{Impact of varying $\beta$ on $\mathrm{SNR}_{\mathrm{th}}$} \label{subsec:varying_beta}
In Section \ref{subsec:bitwise_uep}, we presented an empirical rule for setting $\beta$ as per \eqref{eq:thresh}. Fig. \ref{fig:varying_beta} illustrates the impact of varying $\beta$ for the NR code with $(k,n) = (11,32)$ in an AWGN channel. If $\beta = 0.5 \implies \beta/(1 - \beta) = 1$, then both ACK and NACK are equally protected, hence their BER curves coincide in Fig. \ref{fig:beta_0p5}. As we lower $\beta$ to 0.25, NACK begins to achieve a lower BER than ACK in Fig. \ref{fig:beta_0p25}. However, we observe that the SNR required to achieve a 0.1\% NACK BER is higher than that required to achieve 1\% ACK BER. Finally, at $\beta = 1/11$ in Fig. \ref{fig:beta_0p091}, we observe that the two aforementioned SNRs coincide, which implies that this value of $\beta$ minimizes $\mathrm{SNR}_{\mathrm{th}}$. Furthermore, we observe that $\mathrm{BER}_{\mathrm{NACK}}/\mathrm{BER}_{\mathrm{ACK}} = 0.1$ for a band of SNRs around $\mathrm{SNR}_{\mathrm{th}}$. Similarly, we observe that the BER ratio is $1/3$ and 1 in Fig. \ref{fig:beta_0p25} and \ref{fig:beta_0p5} respectively. We observed this to be true for other values of $p$ as well, hence we arrived at the rule in \eqref{eq:thresh}.
\begin{figure}
\centering
    \subfloat[$\beta = 0.5$]{\includegraphics[width = 2in]{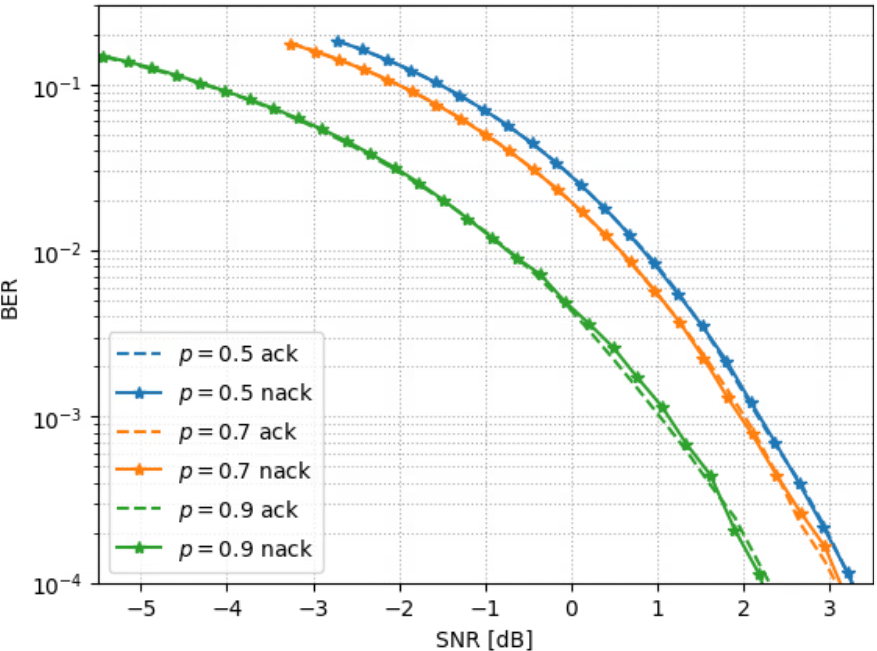}
    \label{fig:beta_0p5}}
    \hspace{0.01in}
    \subfloat[$\beta = 0.25$]{\includegraphics[width = 2in]{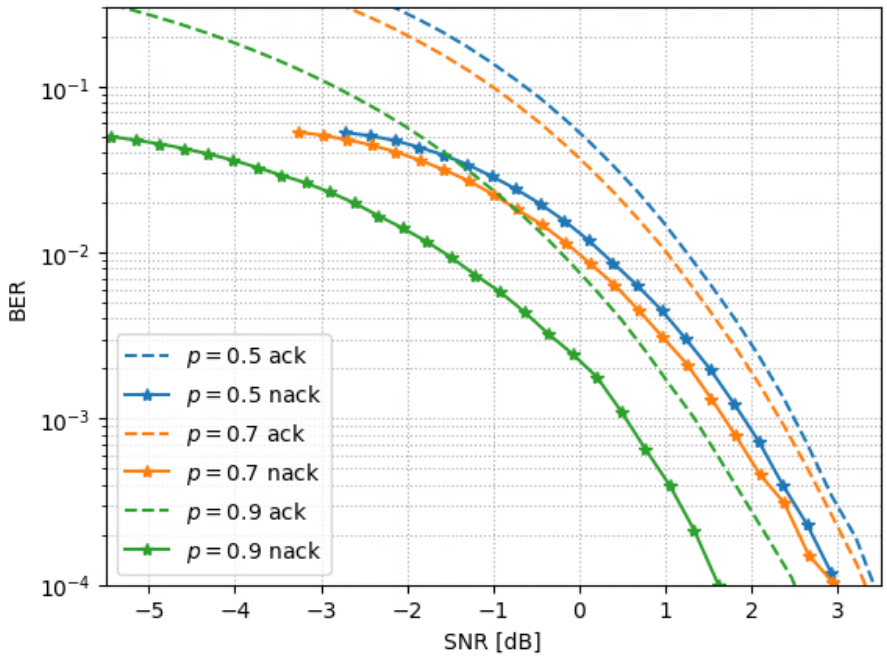}
    \label{fig:beta_0p25}}
    \hspace{0.01in}
    \subfloat[$\beta = 0.091$]{\includegraphics[width=2in]{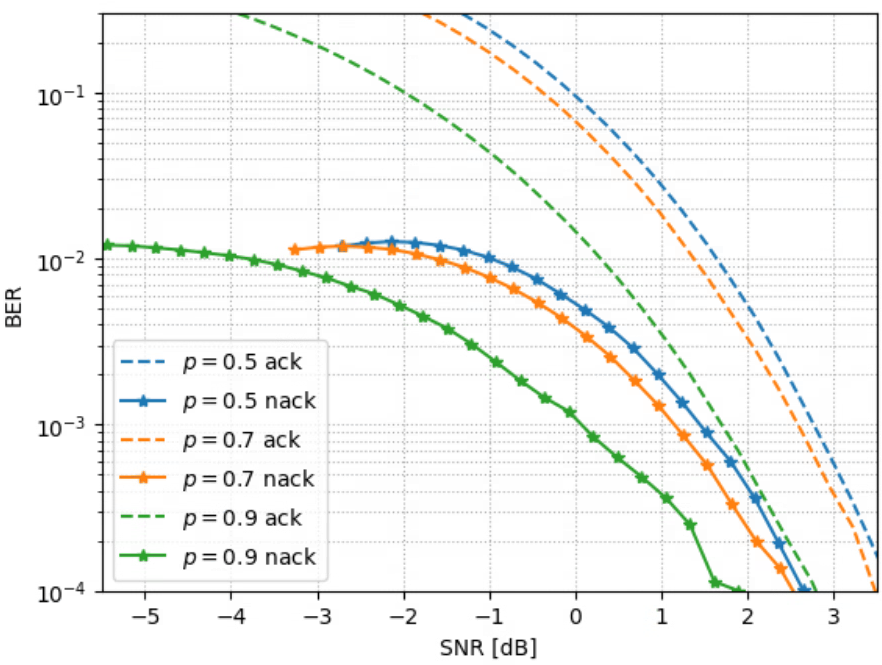}
    \label{fig:beta_0p091}}
\caption{Impact of varying $\beta$ on ACK and NACK BER}
\label{fig:varying_beta}
\end{figure}

\subsection{Derivation of \eqref{eq:Sigma}} \label{subsec:conditional_gaussian}
Note that $\mathbf{\Sigma}_{11} = g^2 \mathbb{E}[\mathbf{h}_{d,l} \mathbf{h}^H_{d,l}]$ ($l$ is the receiver antenna index), from which the formula for $\mathbf{\Sigma}_{11}$ in \eqref{eq:Sigma} follows. To compute $\mathbf{\Sigma}_{12}$, we have
\begin{equation}
\begin{aligned}
    \mathbf{\Sigma}_{12} &= g \mathbb{E}[\mathbf{h}_{d,l} \mathbf{y}^H_{p,l}] \\
    &= g \mathbb{E}[\mathbf{h}_{d,l} (g\sqrt{\beta_m}\mathbf{h}_{p,l}\odot \mathbf{1}_{N_p} + \mathbf{n}_{p,l})^H] \\
    &= g^2 \sqrt{\beta_m} \mathbf{R}_{\mathbf{h}_d \mathbf{h}_p},
\end{aligned}
\end{equation}
since $\mathbf{h}_{d,l}$ and $\mathbf{n}_{p,l}$ are independent and zero mean. Finally, to compute $\mathbf{\Sigma}_{22}$, we have
\begin{equation}
\begin{aligned}
    \mathbf{\Sigma}_{22} &= \mathbb{E}[\mathbf{y}_{p,l} \mathbf{y}^H_{p,l}] \\
    &= \mathbb{E}[(g\sqrt{\beta_m}\mathbf{h}_{p,l}\odot \mathbf{1}_{N_p} + \mathbf{n}_{p,l})(g\sqrt{\beta_m}\mathbf{h}_{p,l}\odot \mathbf{1}_{N_p} + \mathbf{n}_{p,l})^H] \\
    &= g^2 \beta_m \mathbf{R}_{\mathbf{h}_p \mathbf{h}_p} + \sigma^2 \mathbf{I}_{N_p}.
\end{aligned}
\end{equation}

\subsection{Derivation of \eqref{eq:lc_inv_lemma}} \label{subsec:lc_inv_lemma}
We start from the definition of $\mathbf{W}_m$ in \eqref{eq:chest}. Substituting for flat fading and applying the matrix inversion lemma as stated in \eqref{eq:inv_lemma_simple} with $\mathbf{U} = \beta_m \mathrm{SNR} \mathbf{1}_{N_p}$ and $\mathbf{V} = \mathbf{1}_{N_p}^T$, we have
\begin{equation} \label{eq:W}
    \mathbf{W}_m = \mathbf{1}_{N_d}\mathbf{1}_{N_p}^T \beta_m \mathrm{SNR} \bigg(\mathbf{I}_{N_p} - \frac{\mathbf{1}_{N_p}\mathbf{1}_{N_p}^T}{\frac{1}{\beta_m \mathrm{SNR}} + N_p}\bigg).
\end{equation}
Substituting \eqref{eq:W} in the definition of $\mathbf{\Sigma}_{d,l}$ in \eqref{eq:mu_sigma} and simplifying, we obtain
\begin{equation} \label{eq:Sigma_d_l}
    \mathbf{\Sigma}_{d,l} = \mathrm{NV}\bigg(\mathbf{I}_{N_d} + \bigg(\frac{\alpha_m }{\beta_m N_p + \frac{1}{\mathrm{SNR}}}\bigg)\mathbf{c}_m\mathbf{c}_{m}^H \bigg).
\end{equation}
To compute $\mathrm{det} \mathbf{\Sigma}_{d,l}$, we apply $\mathrm{det}$ to \eqref{eq:Sigma_d_l} and utilize the identity $\mathrm{det}(\mathbf{I} + \mathbf{AB}) = \mathrm{det}(\mathbf{I} + \mathbf{BA})$ as
\begin{equation}
    \mathrm{det} \bigg(\mathbf{I}_{N_d} + \bigg(\frac{\alpha_m }{\beta_m N_p + \frac{1}{\mathrm{SNR}}}\bigg)\mathbf{c}_m\mathbf{c}_{m}^H \bigg) = 1 + \frac{\alpha_m N_d}{\beta_m N_p + \frac{1}{\mathrm{SNR}}}
\end{equation}
to obtain the formula in \eqref{eq:lc_inv_lemma}, noting that $\|\mathbf{c}_m\|^2_2 = N_d$. In order to compute $\Delta^2(\mathbf{y}_{d,l},\boldsymbol{\mu}_{d,l})$, we evaluate the inverse of \eqref{eq:Sigma_d_l} using \eqref{eq:inv_lemma_simple} with $\mathbf{U} = \big(\frac{\alpha_m \mathrm{SNR}}{1 + \beta_m N_p \mathrm{SNR}}\big) \mathbf{c}_m$ and $\mathbf{V} = \mathbf{c}_m^H$ to obtain
\begin{equation} \label{eq:inv_Sigma_d_l}
    \mathbf{\Sigma}_{d,l}^{-1} = \frac{1}{\mathrm{NV}} \bigg(\mathbf{I}_{N_d} - \frac{\alpha_m}{\alpha_m N_d + \beta_m N_p + \frac{1}{\mathrm{SNR}}} \mathbf{c}_m \mathbf{c}_m^H \bigg).
\end{equation}
Substituting \eqref{eq:inv_Sigma_d_l} in the definition of $\Delta^2(\mathbf{y}_{d,l},\boldsymbol{\mu}_{d,l})$ in \eqref{eq:mlb_dist}, we obtain its expansion as provided in \eqref{eq:lc_inv_lemma}.

\subsection{Impact of varying ChEST SNR on RMMSE ChEST} \label{subsec:nominal_snr}
\begin{figure}
    \centering
    \includegraphics[width = 3.1in]{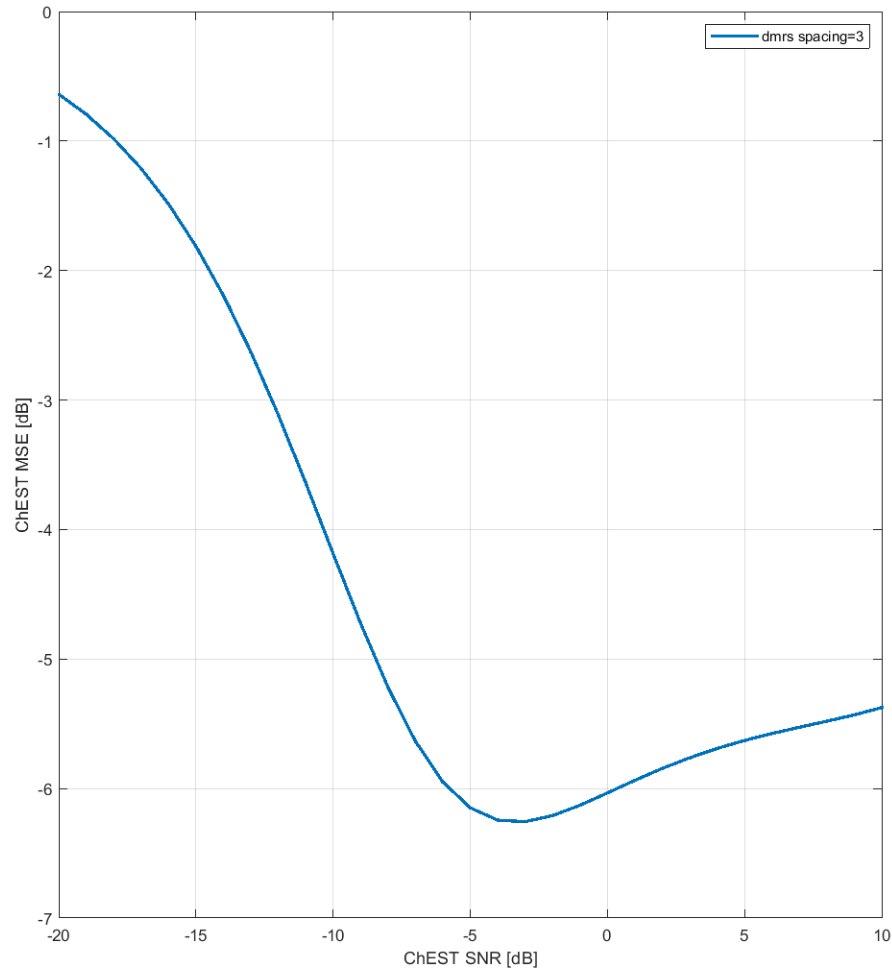}
    \caption{ChEST MSE vs. ChEST SNR. DMRS Spacing of 3 REs and 2 OFDM symbol transmission is assumed.}
    \label{fig:chest_snr}
\end{figure}
In order to understand the impact of varying the nominal SNR, we plot the ChEST MSE as a function of the SNR assumed in ChEST. We fix the channel SNR i.e. the true SNR to $-3$ dB and vary the ChEST SNR from -20 to 10 dB. Assuming RMMSE ChEST, the ChEST MSE can be computed analytically \cite{li2002robust} and is plotted in Fig. \ref{fig:chest_snr}. While underestimating the SNR by 15 dB degrades the MSE by 5 dB, overestimating the SNR by up to 15 dB only causes 0.8 dB of degradation. Hence, we conclude that when the true SNR is low i.e. $\mathrm{SNR} < 0$ dB, there is a negligible downside to overestimating the ChEST SNR. Since PUCCH is typically tested in low SNRs, we set the ChEST SNR $\mathrm{SNR}_n$ to $0$ dB.
\end{document}